\documentclass[a4paper,12pt]{article}%
\usepackage{anysize}
\usepackage{graphicx}
\usepackage{amsmath}
\usepackage{bigints}
\usepackage{amsfonts}
\usepackage{amssymb}
\usepackage{amsthm}
\usepackage{tikz-cd}
\usepackage[all]{xy}
\usepackage{xcolor}
\usepackage{hyperref}
\usepackage{enumerate}
\usepackage{float}
\usepackage[titletoc,title]{appendix}
\providecommand{\U}[1]{\protect\rule{.1in}{.1in}}

\theoremstyle{plain}
\theoremstyle{definition}
\newtheorem{defn}{Definition}[section]

\theoremstyle{remark}

\newtheorem*{post*}{Postulate}

\newtheorem*{rem}{Remark}
\newtheorem{remn}{Remark}

\theoremstyle{remark}
\newtheorem{remnR}{Remark}

\newcommand{\dbar}{\ \ensuremath{\mathchar'26\mkern-12mu d}}

\begin{document}

\title{Finite dimensional thermo-mechanical systems and second order constraints}
\author{Hern\'an Cendra\\{\small Departamento de Matem\'atica} \\{\small Universidad Nacional del Sur, Av. Alem 1253} \\{\small 8000 Bah\'ia Blanca and CONICET, Argentina}
\and Sergio Grillo and Maximiliano Palacios Amaya\\{\small Instituto Balseiro - Comisi\'on Nacional de Energ\'ia At\'omica, Av Bustillo 9.500}\\{\small 8400 San Carlos de Bariloche and CONICET, Argentina}}
\maketitle

\begin{abstract}
In this paper we study a class of physical systems that combine a finite
number of mechanical and thermodynamic observables. We call them \emph{finite
dimensional} \emph{thermo-mechanical systems}. We introduce these systems by
means of simple examples. The evolution equations of the involved observables
are obtained in each example by using, essentially, the Newton's law and the
First Law of Thermodynamics only. We show that such equations are similar to
those defining certain mechanical systems with higher order constraints.
Moreover, we show that all of the given examples can be described in a
variational formalism in terms of \emph{second order constrained systems}.

\end{abstract}
\tableofcontents

\section{Introduction}

%Physical systems are essentially defined by \emph{observables} and laws that govern them. YO PONDR\'{I}A ----- 
Physical systems are essentially defined by
\emph{observables} and \emph{laws}. Typically, the former are magnitudes whose values can be reached by experiments, and the latter are equations that determine and/or relate such values. % (usually equations) which determine and/or relate their possible values. 
If the time is involved in the description of the
system (which can be seen as another observable), the relations between each
observable and the time give precisely the evolution equations of the
system. Roughly speaking, when the observables are positions and
velocities, we say that we have a mechanical system. On the other hand, when
such observables are the temperature, the pressure, the entropy, the volume
and, for instance, the number of moles of certain chemical compounds, we say
that we have a thermodynamic system. In this paper we shall study physical
systems defined, at the same time, by observables of the two mentioned types:
mechanical and thermodynamic observables. In other words, we shall study
physical systems that combine mechanical and thermodynamic \textquotedblleft
degrees of freedom \textquotedblright . We shall call them
\emph{thermo-mechanical systems}. Only those with a finite number of
observables will be considered here.

The paper is divided into two parts. The organization and the content are as
follows. In the first part, in Section \ref{Sec:examples}, we introduce the
idea of thermo-mechanical system by means of several simple examples, i.e. we
give some kind of ostensive definition of such systems. We focus our attention
on finding, for each one of the given examples, the evolution equations of its
corresponding observables. To do that, for the mechanical observables we only
consider the Newton's laws and for the thermodynamic observables we consider
the First Law of Thermodynamics. Regarding the combination of such
observables, we shall assume that the work made on the underlying
thermodynamic system is due to the external (tipically non-conservative)
forces\footnote{By external force we mean any non-conservative force, or a conservative force whose corresponding potential energy is not included in what we consider the mechanical energy of the system.}  acting on the mechanical counterpart. This gives rise to an
\emph{Energy Conservation Principle} which extends those of classical
mechanics and thermodynamics. Such a principle adds another equation to the
set of evolution equations. Among the solutions of the latter, we choose those
for which the evolution of the thermodynamic observables satisfy the Second
Law of Thermodynamics.

Since this kind of systems have not been extensively studied in the
literature,\footnote{Physical systems whose states depend on mechanical and
thermodynamic variables are more common in the domain of continuous media or
fluid dynamics, but not in contexts where only a finite number of degrees of
freedom are involved.} we dedicate some effort on investigating their
evolution equations. By doing that, we conclude that the systems under
consideration can be described as mechanical systems with (higher order)
constraints. This drives us to the second part of the paper, the Section
\ref{Sec:TSaC}, where a precise definition of a thermo-mechanical system is
given inside a variational framework. More precisely, we define such systems
as a particular subclass of the \emph{second order constrained systems} (SOCS)
[see \cite{CG07,G09}]. We describe in this new formalism all the examples
presented in the first part of the paper.

\bigskip

We assume that the reader is familiar with the fundamental concepts of
thermodynamics (see \cite{BBMS,Planck,Fermi,Callen,Zemansky}). However, a
brief review of these concepts can be found in the Appendix. We also assume
some familiarity with basic aspects of classical mechanics (see for instance
\cite{Arnold,Goldstein}). For the second part, a background on Differential
Geometry (see \cite{Boothby,Kobayashi,AMR}) and the ideas of Lagrangian and
Hamiltonian systems in the context of the Geometric Mechanics (see
\cite{Abraham-Marsden,MR}) is expected.\newline

\section{(Examples of) Thermo-mechanical systems}

\label{Sec:examples}

By a \emph{thermo-mechanical system} we shall mean a physical system that
combines both mechanical and thermodynamic degrees of freedom, i.e. a physical
system whose states are defined by mechanical observables, such as positions
and velocities, and thermodynamic observables, e.g. temperature, entropy,
pressure, volume, etc., such that the mentioned observables obey the laws of
classical mechanics and thermodynamics. In the following, in order to grasp an
idea of what kind of systems we are talking about, we introduce some examples
(most of them originally presented in \cite{Palacios}). One of our aims
is to find, for each one of these examples, the equations that determines the
evolution of its corresponding observables.

\subsection{Wagon with internal friction}

\label{wif}

\begin{figure}[h]
\centering
\par
\begin{center}
\includegraphics[scale=0.25]{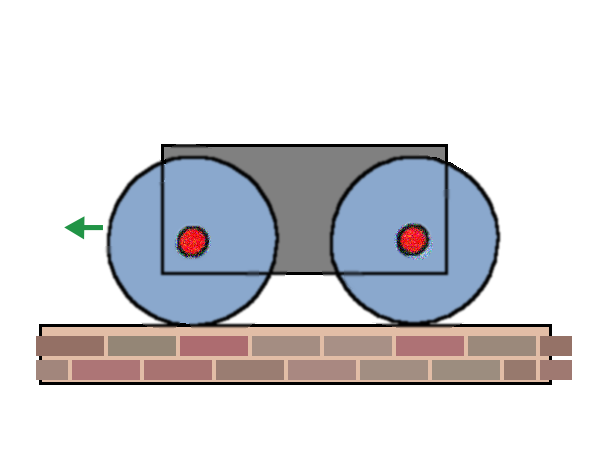}
\end{center}
\caption{Wagon with internal friction}%
\label{G1}%
\end{figure}

Consider a wagon (chassis, axles and wheels) with mass $m$, whose wheels rolls
with no sliding on a horizontal line. The wheels of the wagon are connected by
pairs with an axle, rigidly attached to the wheels, that breaks through the
wagon from side to side (see Figure \ref{G1}). A friction force is present
between each axle and the chassis and it is supposed to be proportional to
their relative angular velocity. 
We shall consider two different situations.

\subsubsection{Thermally isolated wagon}

\label{Ej:CcF}

Assume first that there is no heat exchange between the whole wagon and the
environment. In other words, only adiabatic processes are allowed. The
thermodynamic observables we consider for the system are the internal energy
$U$, the entropy $S$ and the temperature\footnote{The thermal conductivity of
the material, of which the wagon is made, is supposed to be big enough to
ensure that the temperature is well defined and is uniformly distributed for
all time.} $T$. Recall that a review on basic concepts of thermodynamics can be found in the Appendix. We also assume that the only available thermodynamic processes are
the quasi-static ones, in the sense that the state equations hold at every
moment (see Remark \ref{quasi}). We also assume that $U=\nu T$, where $\nu>0$
is the heat capacity of the wagon (which is assumed to be constant). Then,
according to the First Law of Thermodynamics in its infinitesimal form [see
Eq. (\ref{dU})], which in this case reads $dU=T\,dS$, the mentioned observables
must satisfy the following (state) equations at each instant of time%

\begin{equation}
U=\nu T\quad\text{ and }\quad S=\nu\mathrm{ln}\left(  \dfrac{T}{T_{0}}\right)
+S_{0}, \label{Ec:CF EE}%
\end{equation}
where $T_{0}$ and $S_{0}$ are constant.

\begin{remn}
\label{ls1} As we emphasize in the Appendix, every thermodynamic system defines a contact manifold, and its corresponding state equations give rise to a Legendre submanifold of it. In this example, such a manifold is an open subset of $\mathbb{R}^{3}$ 
%$\mathbb{R}^{+}\times\mathbb{R}\times\mathbb{R}$,
with the contact form $\theta:=dU-TdS$ [see Eq. (\ref{tita})], where $(-T,S,U)$ are the global Darboux coordinates, and the Legendre submanifold is defined by Eq. (\ref{Ec:CF EE}). This aspect of thermodynamic systems will be important in the second part of the paper.
\end{remn}

For the mechanical counterpart, we assume that the wheels and the axles have
negligible mass, so the only relevant mechanical observables are the position
$x$ of the chassis and its time derivative $\dot{x}$. The Newton's laws say
that%
\begin{equation}
m\ddot{x}=-\mu\dot{x}, \label{AH:Newq}%
\end{equation}
being $\mu>0$ a constant depending on the friction torque between the axle and
the chassis (and also on the diameter of the wheels). Its general solution is%
\begin{equation}
x(t)=-\dfrac{\dot{x}_{0}m}{\mu}\left(  e^{\frac{-\mu t}{m}}-1\right)  +x_{0}.
\label{xt}%
\end{equation}
This says that the internal friction of the axles will slow down the wagon
exponentially, as expected.

Consider again the First Law of Thermodynamics [see Eq. (\ref{deltaU})]. It
says that the variation $\Delta U$ of the internal energy along any process is
equal to the heat $\mathbb{Q}$ exchanged between the wagon and the environment
(along such a process) minus the mechanical work $W$ made on the wagon, i.e.
$\Delta U=\mathbb{Q}-W$. On one hand, by the adiabaticity condition assumed
above, we know that $\mathbb{Q}=0$. On the other hand, the unique forces that
make work are the friction forces between the axles and the chassis (the force
exerted on the wheels by the floor, which allows the wheels to roll with no
sliding, does not make work), and such a work is equal to $\Delta E_{mec}$,
i.e. the variation of the mechanical energy $E_{mec}=m\dot{x}^{2}/2$.

\begin{remn}\label{ExtForce}
As usual, a friction force acting on a mechanical system (see the right hand side of Eq. (\ref{AH:Newq})) is conceived as an external force. Also, this kind of forces are non-conservative. This is why the mechanical energy $E_{mec}$ is given by the kinetic energy only.
\end{remn}

Thus,
$W=\Delta E_{mec}$, or equivalently $\Delta E_{mec}+\Delta U=0$. Consequently,
if we define the \emph{total energy }of the system as
\begin{equation}
E:=E_{mec}+U=\frac{m\dot{x}^{2}}{2}+\nu T, \label{E}%
\end{equation}
then $E$ is conserved for all $t$, namely [see (\ref{AH:Newq})], the identity%
\begin{equation}
\dfrac{d}{dt}\left(  \dfrac{m\dot{x}^{2}}{2}+\nu T\right)  =m\ddot{x}\dot
{x}+\nu\dot{T}=-\mu\dot{x}^{2}+\nu\dot{T}=0 \label{cec}%
\end{equation}
must hold. As a consequence $\dot{T}=\dfrac{\mu\dot{x}^{2}}{\nu}$, and from
(\ref{xt}) we have that%
\begin{equation}
T\left(  t\right)  =-\frac{m\dot{x}_{0}^{2}}{2\nu}\,e^{\frac{-2\mu t}{m}%
}+T_{0}. \label{tdt}%
\end{equation}
Therefore, using (\ref{Ec:CF EE}) and the last equation we have, for each initial
condition, only one thermodynamic process $t\longmapsto\left(  T\left(
t\right)  ,S\left(  t\right)  ,U\left(  t\right)  \right)  $. Note that,
since
\[
\dot{S}=\dfrac{\nu\dot{T}}{T}=\dfrac{\mu\dot{x}^{2}}{T},
\]
the entropy increases in time when the wagon moves, i.e. the Second Law of
Thermodynamic holds for all of these processes. Moreover, such processes are
(generically) irreversible. In fact, in an isolated system as it stands, a
thermodynamic process is reversible if only if the change of entropy vanishes.
In our example, this is only possible if the velocity of the wagon is zero.

\bigskip

Summing up, we have a thermo-mechanical system defined by the observables $x$,
$\dot{x}$, $T$, $S$ and $U$, whose evolution equations are given by Equations
(\ref{Ec:CF EE}), involving the thermodynamic observables, Eq. (\ref{AH:Newq}%
), involving the mechanical ones, and Eq. (\ref{cec}), linking both of them.
All of these equations enable us, given an initial condition, to find a unique
temporal evolution $t\longmapsto\left(  x\left(  t\right)  ,\dot{x}\left(
t\right)  ,T\left(  t\right)  ,S\left(  t\right)  ,U\left(  t\right)  \right)
$.

\subsubsection{The wagon in a thermal bath}

\label{wter}

Assume now that heat exchange by conduction, between the wagon and the
environment, is permitted, and that the wagon is immersed in a thermal bath of
(constant) temperature $T_{b}$. The observables of the system $x$, $\dot{x}$,
$T$, $S$ and $U$ will be again subjected to the Eqs. (\ref{Ec:CF EE}) and
(\ref{AH:Newq}). But now the total energy [see Eq. (\ref{E})] is not
conserved. In fact, as we discussed above, since $\Delta U=\mathbb{Q}-W$ and
$\Delta E_{mec}=W$, with $W$ given by the friction forces, we have that
$\Delta E=\Delta E_{mec}+\Delta U=\mathbb{Q}$, or in infinitesimal terms
$dE= \dbar\mathbb{Q}$ (see Eq. (\ref{Eq:ConsU}) in the Appendix).

If no further information is available for $\dbar\mathbb{Q}$, we can only say that the temporal evolution of the system is given by curves $t\longmapsto\left(  x\left(  t\right)  ,\dot
{x}\left(  t\right)  ,T\left(  t\right)  ,S\left(  t\right)  ,U\left(
t\right)  \right) $ satisfying (\ref{Ec:CF EE}), (\ref{AH:Newq}) and the
Second Law of Thermodynamics [see Eq. (\ref{sl3})], which in this case says
that $dS\geq dE/T$.

Assume that the \emph{Fourier's law} holds, i.e. the heat exchanged by
conduction along a time $dt$, between a body at temperature $T$ and a thermal
bath at temperature $T_{b}$, is given by $\dbar\mathbb{Q}=\kappa\,\mathcal{A}\,\left(  T_{b}-T\right)  \,dt$.
Here $\kappa>0$ is the conduction coefficient (assumed constant) and
$\mathcal{A}$ is related to the area though which the heat flux takes place.
For simplicity, we shall take $\mathcal{A}$ equal to $1$. Then $\dot
{E}=-\kappa\,\left(  T-T_{b}\right)  $, i.e. [compare to Eq. (\ref{cec})]
\begin{equation}
-\mu\dot{x}^{2}+\nu\dot{T}=-\kappa\,\left(  T-T_{b}\right)  .\label{nce}%
\end{equation}
As a consequence, under such assumptions, the evolution equations are
(\ref{Ec:CF EE}), (\ref{AH:Newq}) and (\ref{nce}), and they determinate
completely the evolution of all the observables. The solutions to
(\ref{AH:Newq}) and (\ref{nce}) are given by (\ref{xt}) and%
\begin{equation}
T\left(  t\right)  =-\left(  T_{b}-T_{0}-\dfrac{m\dot{x}_{0}^{2}}{2\nu
}\right)  \,e^{-\frac{\kappa}{\nu}t}-\dfrac{m\dot{x}_{0}^{2}}{2\nu}%
\,e^{-\frac{2\mu}{m}t}+T_{b},\label{tt}%
\end{equation}
respectively. Then $S(t)$ and $U(t)$ can be constructed by combining (\ref{Ec:CF EE})
with (\ref{tt}). Note that, according to (\ref{nce}),
\[
dS=\dfrac{\nu\,\dot{T}}{T}\,dt=\frac{\mu\,\dot{x}^{2}}{T}\,dt+\kappa\,\left(
\frac{T_{b}}{T}-1\right)  \,dt\geq\kappa\,\left(  \frac{T_{b}}{T}-1\right)
\,dt=\frac{\dbar\mathbb{Q}}{T},
\]
what implies that each curve $t\longmapsto\left(  T\left(  t\right)  ,S\left(
t\right)  ,U\left(  t\right)  \right)  $ satisfies the Second Law and defines
an irreversible process (at least when the wagon moves). In particular, above
assumptions, including the Fourier's law, are compatible with the Second Law
of Thermodynamics.

\begin{rem}
Eqs. (\ref{Ec:CF EE}), (\ref{AH:Newq}) and (\ref{cec}) [resp. (\ref{nce})]
define a system of differential-algebraic equations, as those appearing in
mechanical systems with constraints. This observation will be further
exploited in Section \ref{Sec:TSaC}.
\end{rem}

\subsection{Vertical piston}

\label{Ej:EVA}

\begin{figure}[h]
\begin{center}
\centering
\includegraphics[scale=0.45]{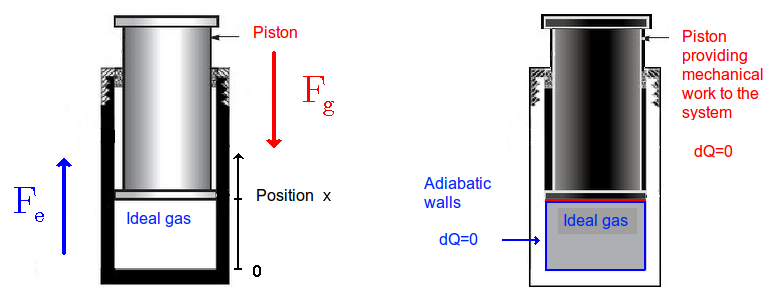}
\end{center}
\caption{Mechanical and thermodynamical scheme of the piston}%
\label{G2}%
\end{figure}

Consider an ideal gas (made of one chemical compound) confined in a cylinder
by a vertical piston of mass $m$ (see Figure \ref{G2}). We shall consider two different kind of allowed processes.

\subsubsection{Adiabatic processes}

\label{adcase}

Assume that the piston and the cylinder, i.e. the container of the gas, are
perfect thermal insulators and that there is no friction between them. The
only mechanical observables are the piston's position $x$ and velocity
$\dot{x}$. Since the number of moles $N$ of the gas is constant, say
$N:=N_{0}$, the relevant thermodynamic extensive variables are the internal
energy $U$, the volume $V$ and the entropy $S$, and the intensive ones are the
pressure $P$ and the temperature $T$ of the ideal gas.
%(we can omit the chemical potential).
It can be shown that these observables are related by the equations [see
(\ref{EEIG})]%
\begin{equation}
PV=N_{0}RT, \label{Ec:GI1}%
\end{equation}%
\begin{equation}
U=\alpha N_{0}RT \label{Ec:GI2}%
\end{equation}
and%
\begin{equation}
S=S_{0}+N_{0}R\ \mathrm{ln}\left(  \dfrac{T^{\alpha}V}{T_{0}^{\alpha}V_{0}%
}\right)  , \label{Ec:GI3}%
\end{equation}
where $S_{0}$, $V_{0}$ and $T_{0}$ are constant with units of entropy, volume
and temperature, respectively. Since the container of the gas is a thermal
insulator, only adiabatic processes are allowed. As a consequence, the
identity%
\begin{equation}
PV^{\gamma}=k \label{Ec:GIA}%
\end{equation}
also holds, where $k$ is a constant and $\gamma=\frac{\alpha+1}{\alpha}$.
Then, Equations (\ref{Ec:GI1}), (\ref{Ec:GI3}) and (\ref{Ec:GIA}) imply that
all processes must be isoentropic (as it is well known for the ideal gas),
i.e.%
\begin{equation}
S=S_{0}, \label{Scero}%
\end{equation}
and consequently reversible. In summary, the thermodynamic observables must
fulfill the Eqs. (\ref{Ec:GI1}), (\ref{Ec:GI2}), (\ref{Ec:GIA}) and
(\ref{Scero}). Note that the Second Law is automatically satisfied.

\begin{remn}
\label{ls2} For further convenience, and regarding the contact structure related to every thermodynamic system (see the Appendix), let us mention that the equations (\ref{Ec:GI1}) (\ref{Ec:GI2}) and
(\ref{Ec:GI3}) define a Legendre submanifold $\mathcal{N}\subset\mathbb{R}%
^{5}$, w.r.t. the contact form $\theta:=dU-T\ dS+\,P\,dV$ [see Eq. (\ref{tita})], which encodes the
equilibrium states of the ideal gas. The additional equation (\ref{Scero})
gives a submanifold of $\mathcal{N}$ where the adiabatic processes of the
ideal gas are contained.
\end{remn}

On the mechanical side, from Newton's laws, the position $x$ of the piston
must satisfy
\begin{equation}
m\ddot{x}=\underset{F_g}{\underbrace{-mg}}+\underset{F_e}{\underbrace{PA}}, \label{Ec:New EVA}%
\end{equation}
where $A$ is the area of the horizontal section of the piston and $g$ is the
acceleration of gravity, so the force associated is labeled $F_g$. Note that the force made by the pressure is an external force, from the point of view of the mechanical degrees of freedom, so we label it $F_e$. Because of the geometric configuration, we have that
\begin{equation}
Ax=V, \label{Ec:EA VinG}%
\end{equation}
then, using Eq. (\ref{Ec:GIA}), follows that
\begin{equation}
P=\dfrac{k}{(Ax)^{\gamma}}. \label{Pdeq}%
\end{equation}
As a consequence,%

\begin{equation}
m\ddot{x}=F_g+F_e=-mg+\dfrac{kA}{(Ax)^{\gamma}}. \label{Ec:EA Newton}%
\end{equation}
The solutions of the last equation are given by the quadrature%

\begin{equation}
\displaystyle\bigintss_{1}^{x\left(  t\right)  }\dfrac{\sigma}{\sqrt
{c_{1}+2\left(  -gs+\dfrac{kA}{mA^{\gamma}(1-\gamma)}s^{1-\gamma}\right)  }%
}ds=t+c_{2}. \label{Ec:EA Sol Cuad}%
\end{equation}
Here, $c_{1}$ and $c_{2}$ are integration constants, and $\sigma=\pm1$
(depending on the initial conditions). On the other hand, plugging the
identity $Ax=V$ on Equations (\ref{Ec:GI1}) and (\ref{Ec:GI2}), we have that%
\begin{equation}
T=\dfrac{k\left(  Ax\right)  ^{1-\gamma}}{N_{0}R}\ \ \ \text{and\ \ \ }%
U=\alpha k\left(  Ax\right)  ^{1-\gamma}. \label{Tdeq}%
\end{equation}
Thus, as in the previous example, for each initial value of the mechanical
observables we have only one (adiabatic) thermodynamic process, defined by a
time-parametrized curve
\begin{equation}
t\longmapsto\left(  P\left(  t\right)  ,T\left(  t\right)  ,V\left(  t\right)
,S\left(  t\right)  ,U\left(  t\right)  \right)  , \label{curve}%
\end{equation}
with [see Eqs. (\ref{Scero}), (\ref{Pdeq}) and (\ref{Tdeq})]%

\begin{equation}%
\begin{cases}
P(t)=k\,\left(  A\,x(t)\right)  ^{-\gamma},\\
T(t)=\dfrac{k\,\left(  A\,x(t)\right)  ^{1-\gamma}}{N_{0}\,R},\\
V(t)=A\,x(t),\\
S(t)=S_{0},\\
U\left(  t\right)  =\alpha\,k\,\left(  A\,x\left(  t\right)  \right)
^{1-\gamma},
\end{cases}
\label{Ec:EA Sol}%
\end{equation}
and where $x(t)$ is given by the Eq. (\ref{Ec:EA Sol Cuad}). As we said above,
since the process is adiabatic and isoentropic, then it is reversible.

In conclusion, this example consists in a thermo-mechanical system defined by
observables $x$, $\dot{x}$, $P$, $T$, $V$, $S$ and $U$, with evolution
equations given by (\ref{Ec:EA Newton}), mainly related to the mechanical
degrees of freedom, the Eqs. (\ref{Ec:GI1}), (\ref{Ec:GI2}), (\ref{Ec:GIA})
and (\ref{Scero}), exclusively related to the thermodynamic ones, and the Eq.
(\ref{Ec:EA VinG}) that links both of them. In the next subsection we shall
analyze a little bit closer the solutions of these equations.

\subsubsection{Reversibility and quasi-staticity}

\label{revq}

It is easy to show that Eq. (\ref{Ec:EA Newton}) is equivalent to the system
of first order ordinary differential equations%

\[%
\begin{cases}
\dot{q}=\dfrac{p}{m},\\
\dot{p}=-mg+kA^{-\gamma+1}q^{-\gamma}.
\end{cases}
\]
Such a system defines a Hamiltonian vector field with Hamiltonian function
\begin{equation}
H(q,p):=\dfrac{p^{2}}{2m}+mgq+\dfrac{\alpha k}{A^{\gamma-1}}q^{1-\gamma}.
\label{hu}%
\end{equation}

\begin{figure}[h]
\centering
\parbox{2.5in}{\includegraphics[scale=0.4]{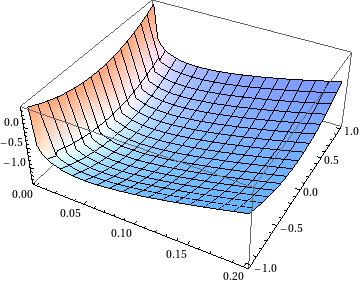}
\caption{Graph of $H$}\label{fig:2figsA}}\qquad\qquad\begin{minipage}{2in}%
\includegraphics[scale=0.35]{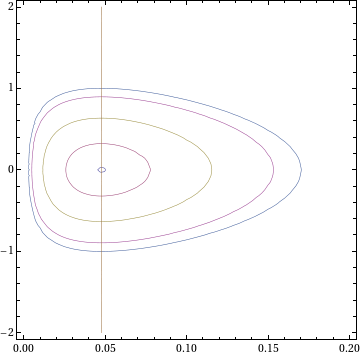}
\caption{Orbits}%
\label{fig:2figsB}%
\end{minipage}\end{figure}

As it is well known, the Hamiltonian is conserved along the solutions. This
conservation property will be analyzed in Section \ref{enco}. Regarding its
stability, it can be shown that the only equilibrium point is
\[
(q^{\ast},p^{\ast})=\left(  \dfrac{kA^{1-\gamma}}{mg},0\right)  ,
\]
which is a center. In the neighborhood of such center the solutions are closed
orbits. For instance, when $m=1$, $\,g=1$, $A=1$, $\alpha=\frac{3}{2}$,
$\gamma=\frac{5}{3}$ and $k=1$, we have the behavior shown in Figures
$\ref{fig:2figsA}$ and $\ref{fig:2figsB}$. As a consequence, the solutions $x$
of Eq. (\ref{Ec:EA Newton}) are oscillatory. This implies that all the
observables of the system have an oscillatory behavior [recall Eq.
(\ref{Ec:EA Sol})], which is consistence with the fact that, as we said
before, each thermodynamic process (one for each initial condition) defined by
(\ref{curve}) is reversible. Also, as it is known, near the equilibrium,
speeds are slow. Then, around such a point, the quasi-static condition is
satisfied in the usual sense (see Remark \ref{quasi}).

\subsubsection{Isothermal processes}

\label{isot}

Assume now that, instead of the adiabaticity condition (\ref{Ec:GIA}), we ask
the temperature of the gas to be constant. Thus, the observables of the system
$x$, $\dot{x}$, $P$, $T$, $V$, $S$ and $U$ satisfy again the Equations
(\ref{Ec:GI1}), (\ref{Ec:GI2}), (\ref{Ec:GI3}), (\ref{Ec:EA Newton}) and
(\ref{Ec:EA VinG}), but (\ref{Ec:GIA}) [or equivalently (\ref{Scero})] must be
replaced by the condition $T=T_{0}$. In this case, Eq. (\ref{Ec:EA Newton})
translates to%
\begin{equation}
m\ddot{x}=-mg+\dfrac{N_{0}RT_{0}}{x}.\label{nei}%
\end{equation}
Solving this equation, we have for the rest of the observables the following
expressions:%
\begin{equation}%
\begin{cases}
P(t)=\frac{N_{0}RT_{0}}{Ax\left(  t\right)  },\\
T(t)=T_{0},\\
V(t)=A\,x(t),\\
S(t)=S_{0}+N_{0}R\ \mathrm{ln}\left(  x\left(  t\right)  /x_{0}\right)  ,\\
U\left(  t\right)  =\alpha N_{0}RT_{0}\,,
\end{cases}
\label{rei}%
\end{equation}
being $x_{0}$ the initial position of the piston. 

\begin{remn}
\label{rever}As in the adiabatic case, since the solutions of Eqs. (\ref{nei})
are invariant by changing $t$ by $-t$, then the curves given by Eqs.
(\ref{rei}) define reversible processes.
\end{remn}

\subsubsection{The total energy conservation}

\label{enco}

Let us go back to Section \ref{adcase} (the adiabatic case of the vertical
piston). It can be shown by direct calculations that the quantity [compare to
Eq. (\ref{hu})]
\begin{equation}
E:=E_{mec}+U:=\frac{m\dot{x}^{2}}{2}+mgx+U \label{te}%
\end{equation}
[compare to Eq. (\ref{E})], that we can call the \emph{total energy} of the
system, is conserved along the curves that solve the Eqs. (\ref{Ec:GI1}),
(\ref{Ec:GI2}), (\ref{Ec:GIA}), (\ref{Scero}), (\ref{Ec:EA Newton}) and
(\ref{Ec:EA VinG}). 

\begin{rem}
Note that we are considering that $E_{mec}$ is constituted by, besides the kinetic energy, just the potential energy related to $F_g=-mg$. We are not including the potential energy related to $F_e=kA^{-\gamma+1}x^{-\gamma}$ because, as previously mentioned, we are taking it as an external force.
\end{rem}

Moreover, it is easy to show that the last system of
equations is equivalent (modulo an additive integration constant) to the
systems given by Eqs. (\ref{Ec:GI1}), (\ref{Ec:GI2}), (\ref{Ec:GIA}),
(\ref{Scero}), (\ref{Ec:EA Newton}) and the condition $\dot{E}=0$, i.e.%
\begin{equation}
m\,\ddot{x}\,\dot{x}+mg\,\dot{x}+\dot{U}=0. \label{sicon}%
\end{equation}
That is to say, the equation given by the geometrical constraint between the
position $x$ and the volume $V$ can be replaced by an energy conservation
condition. Let us analyze why this conservation holds.

On one hand, the First Law of Thermodynamics says that $\Delta U=\mathbb{Q}-W$
[see Eq. (\ref{deltaU})]. In the system under consideration, $\mathbb{Q}=0$
and $W$ is the work made by the pressure of the gas. In particular $\Delta
U=-W$. On the other hand, by the laws of classical mechanics, the variation
$\Delta E_{mec}$ of the mechanical energy is equal to the work made by the
external forces acting on the underlying mechanical system, which in this case
(see previous Remark) is given by the pressure of the gas. Then $\Delta E_{mec}=W$, and consequently
$\Delta E_{mec}+\Delta U=0$. So the total energy conservation property is due
to the fact that:

\begin{description}
\item[\textbf{W}] \emph{the work made on the underlying thermodynamic system
is precisely the work made by the external forces acting on the underlying
mechanical counterpart}.
\end{description}

In other words, the internal energy $U$ is defined by the work of the
%non-conservative
external forces appearing in the underlying mechanical system (see Remark
\ref{work}). Notice that the same is true for the wagon with internal friction (see Remark \ref{ExtForce} and the paragraph before it).

\bigskip

Now, let us go back to Section \ref{isot}. If we calculate the derivative of
the total energy $E$ [see Eq. (\ref{te})] along the curves solving Eqs. (\ref{nei})
and (\ref{rei}), we obtain that $\dot{E}=T_{0}\,\dot{S}$, i.e.%
\begin{equation}
m\,\ddot{x}\,\dot{x}+mg\,\dot{x}+\dot{U}=T_{0}\,\dot{S}.\label{noncon}%
\end{equation}
In particular, the total energy is not conserved in this case. Let us mention
that, contrary to what happens for the adiabatic situation, the geometrical
condition in Eq. (\ref{Ec:EA VinG}) can not be replaced by Eq. (\ref{noncon}).

\begin{rem}
Since the statement $\mathbf{W}$ holds in this case too, we have that $\Delta
U=\mathbb{Q}-W$ and $\Delta E_{mec}=W$, and since $\Delta U=0$ (because
$U=\alpha N_{0}RT_{0}=cte$), then $\Delta E=\Delta E_{mec}=\mathbb{Q}$. So,
from the identity $\dot{E}=T_{0}\,\dot{S}$, or equivalently $\Delta
E=T_{0}\,\Delta S$, we have that $\Delta S=\mathbb{Q}/T_{0}$. This implies
that all the involved processes satisfy the Second Law and are reversible, as
mentioned in Remark \ref{rever} [see also Eq. (\ref{Eq:QE})].
\end{rem}

We shall assume form now on that statement $\mathbf{W}$ holds for all the
thermo-mechanical systems we consider in this paper. This is clearly
equivalent to the following extension of the\emph{ Energy Conservation
Principle}:

\begin{description}
\item[\textbf{ECP}] \emph{Consider a thermo-mechanical system whose underlying
mechanical (resp. thermodynamic) system has energy }$E_{mec}$\emph{ (resp.
}$U$\emph{). If, along a given process, the heat }$\mathbb{Q}$\emph{ exchanged
between the underlying thermodynamic system and the environment is zero, then
the total energy }$E:=E_{mec}+U$\emph{ is preserved, i.e. }$\Delta
E_{mec}+\Delta U=0$\emph{. In general, we shall have that }$\Delta
E_{mec}+\Delta U=\mathbb{Q}$\emph{.}
\end{description}

\subsection{Dissipative vertical piston}

\label{Sec:EVDF}

Consider again the gas, the cylinder and the piston, but this time suppose
that there exists friction between the cylinder and the piston. In addition,
suppose that dissipation by conduction is present between the gas and its
container (cylinder and piston). As in Section \ref{adcase}, we assume that no
heat exchange takes place between the container and the environment. We model
the mentioned friction in the same way as we did for the wagon, where the
massless cylinder and the massive piston would play the role of the massless
axles and the massive chassis, respectively. Shortly speaking, we are
considering a sort of combination of the main features of the
thermo-mechanical systems studied in Sections \ref{wif} and \ref{Ej:EVA}.

The observables in this case are:

\begin{itemize}
\item[a)] the position $x$ and velocity $\dot{x}$ of the piston;

\item[b)] the internal energy $U_{c}$, the temperature $T_{c}$ and the entropy
$S_{c}$ of the whole container (i.e. the cylinder and the piston), for which
we assume (as we did for the wagon) that $U_{c}=\nu T_{c}$;

\item[c)] the internal energy $U$, the pressure $P$, the volume $V$, the
temperature $T$ and the entropy $S$ of the ideal gas.
\end{itemize}

Note that, with respect to the last example, new thermodynamic variables are
present. Moreover, we can see this new example as a composite system (see
Remark \ref{comp}), with subsystems given by the \textquotedblleft
container\textquotedblright\ and the \textquotedblleft gas\textquotedblright.
In particular, the internal energy of the whole underlying thermodynamic
system is $U+U_{c}=U_{tot}$.

Allowing again only quasi-static processes, the thermodynamic observables must
satisfy, at every time, the state equations%
\begin{equation}
U_{c}=\nu T_{c}\ \ \ \text{and}\ \ \ S_{c}=\nu\mathrm{ln}\left(  \dfrac{T_{c}%
}{T_{c,0}}\right)  +S_{c,0} \label{eec}%
\end{equation}
for the container, and
\begin{equation}
PV=N_{0}RT,\ \ \ U=\alpha N_{0}RT\ \ \ \text{and}\ \ \ S=S_{0}+N_{0}%
R\ \mathrm{ln}\left(  \dfrac{T^{\alpha}V}{T_{0}^{\alpha}V_{0}}\right)
\label{eeg}%
\end{equation}
for the gas. On the mechanical side, we have that%

\begin{equation}
m\ddot{x}=-mg+PA-\mu\dot{x}, \label{Ec:ECF1}%
\end{equation}
where $\mu$ is the friction coefficient between the cylinder and the piston,
which we are assuming constant. Recalling that
\begin{equation}
V=Ax \label{gc}%
\end{equation}
and using the first formula of Eq. (\ref{eeg}), we have that
\[
PA=\frac{N_{0}RT}{x},
\]
so Eq. (\ref{Ec:ECF1}) translates to%
\begin{equation}
m\ddot{x}=-mg+\frac{N_{0}RT}{x}-\mu\dot{x}. \label{enm}%
\end{equation}
On the other hand, since there is no heat exchange between the whole system
(container plus gas) and the environment, assumption $\mathbf{ECP}$ (see at
the end of Section \ref{enco}) says that the total energy%
\[
E=E_{mec}+U+U_{c}=\frac{m\dot{x}^{2}}{2}+mgx+\alpha N_{0}RT+\nu T_{c}%
\]
must be conserved, i.e.%
\begin{equation}
\left(  m\ddot{x}+mg\right)  \,\dot{x}+\alpha N_{0}R\dot{T}+\nu\dot{T}_{c}=0.
\label{ce}%
\end{equation}

As for the wagon in a thermal bath, if no further information is given for
heat transferred per time unit, we can not determine completely the
evolution. We can only say that it is given by curves satisfying the Equations
(\ref{eec}), (\ref{eeg}), (\ref{gc}), (\ref{enm}) and (\ref{ce}), and the
Second Law. 

Assume again that the Fourier's law holds. Let us consider the subsystem
formed out by the container (cylinder and piston). According to such law (as seen in Section \ref{wter}),
the variation $\dot{E}_{mec}+\dot{U}_{c}$ of its total energy per time unit
must be equal to $-\kappa\,\mathcal{A}\,\left(  T-T_{c}\right)  $ (the heat
transferred per time unit), that is to say%
\begin{equation}
\left(  m\ddot{x}+mg\right)  \,\dot{x}+\nu\dot{T}_{c}=\kappa\,\mathcal{A}%
\,\left(  T-T_{c}\right)  ,\label{Fl}%
\end{equation}
where $\kappa$ is the conduction coefficient (which we assume constant) and
$\mathcal{A}$ is the area through which heat flows, which is a linear function
on $x$. Combining the last equation with (\ref{ce}), it follows that%

\begin{equation}
\dot{T}=\frac{\kappa\mathcal{A}}{\alpha N_{0}R}\,\left(  T_{c}-T\right)
.\label{Ec:ECF2}%
\end{equation}
The system of ordinary differential equations given by Eq. (\ref{enm}),
(\ref{ce}) and (\ref{Ec:ECF2}) determine completely the evolution of the
observables $x$, $T$ and $T_{c}$. It is clear that the evolution of the rest
of observables can be derived from Eqs. (\ref{eec}), (\ref{eeg}) and (\ref{gc}).

\bigskip

\begin{figure}[h]
\centering
\par
\begin{center}
\includegraphics[scale=0.4]{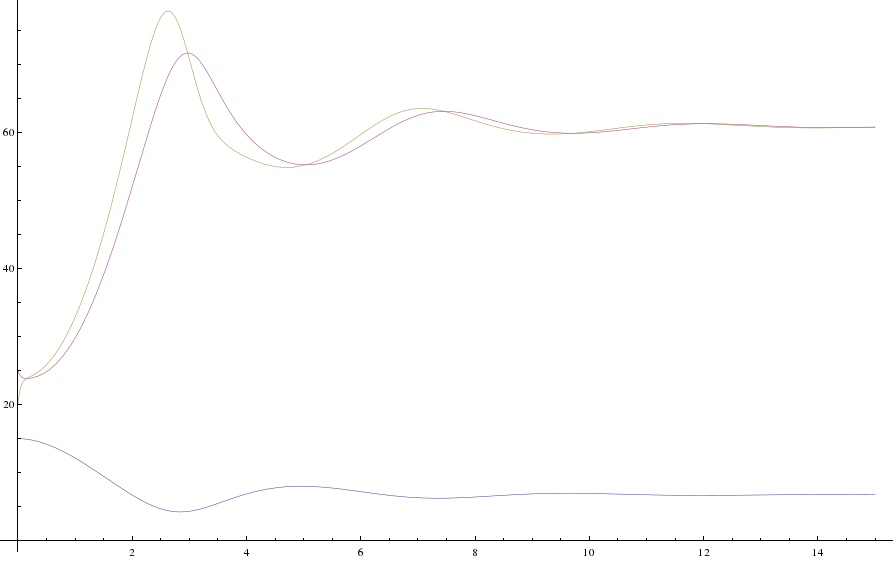}
\end{center}
\caption{Temperatures and position of the dissipative piston with friction}%
\label{G3}%
\end{figure}

\begin{figure}[h]
\centering
\par
\begin{center}
\includegraphics[scale=0.4]{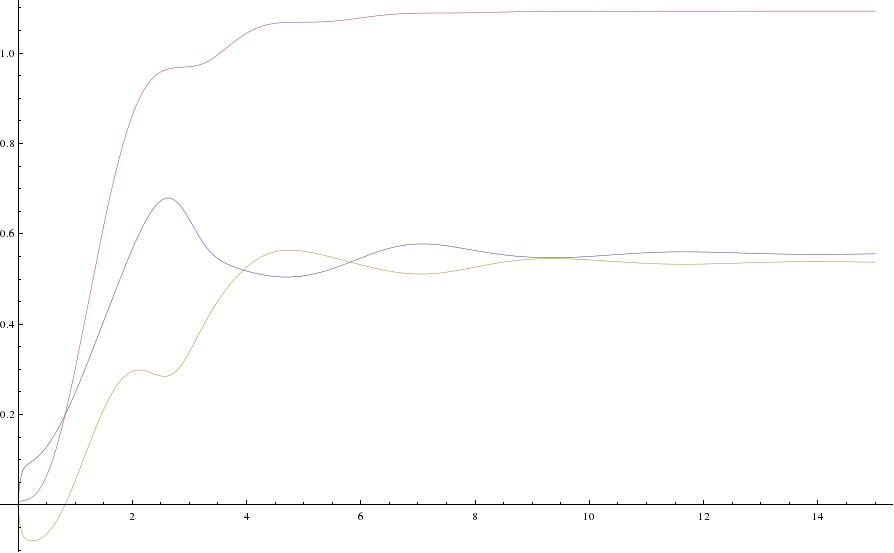}
\end{center}
\caption{Entropies of the dissipative piston with friction}%
\label{G4}%
\end{figure}

To visualize the behavior of this system [see Figures \ref{G3} and \ref{G4}],
we solve numerically\footnote{Using NDSolve in Wolfram Mathematica.} the Equations (\ref{Ec:ECF2}), (\ref{enm}) and (\ref{ce}) (for the
unknowns $T$, $T_{c}$ and $x$) in the case in which $g=9$, $m=1$, $\kappa
=0.2$, $\mu=0.8$, $\nu=0.5$, $N_{0}R=1$, $\alpha=3/2$ and with initial
conditions $x(0)=15$, $\dot{x}(0)=0$, $T(0)=25$ and $T_{c}(0)=20$. In Figure
\ref{G3}, the red curve is the temperature of the ideal gas, the blue one is
the position of the piston, and the yellow one is the temperature of the
container.\newline

The evolution of the entropies is shown in Figure \ref{G4}, where we
are setting $S_{0}=0=S_{c,0}$. The red curve is the total entropy
$S(t)+S_{c}(t)$, the blue one is $S(t)$ and the yellow is $S_{c}(t)$. In such
a figure, it can be seen the monotonous behavior of the total entropy, which
agrees with the statement of the Second Law of Thermodynamics.

\begin{rem}
It might happen, for another choice of parameters, that some curves do not satisfy the Second Law. In such a
case, we must select those curves that do fulfill such law.
\end{rem}

To summarize, this thermo-mechanical system is defined by observables $x$,
$T_{c}$, $S_{c}$, $U_{c}$, $P$, $T$, $V$, $S$ and $U$, subjected to the Eq.
(\ref{Ec:ECF1}) for the mechanical side, the Eqs. (\ref{eec}), (\ref{eeg}) and
(\ref{Ec:ECF2}) for the thermodynamic counterpart, and Eqs. (\ref{gc}) and
(\ref{ce}) coupling the previous ones.

\begin{rem}
It is well known that Fourier's law of heat conduction is empirical, so unlike
previous examples where the validity of the Second Law of Thermodynamics may
be inferred by construction or simply calculations, in this example it is not
obvious that the total entropy of the system increases in time; fact shown in
Figure \ref{G4}.
\end{rem}

\section{Thermo-mechanical systems and constraints}

\label{Sec:TSaC}

We saw in the previous section that the observables of a thermo-mechanical
system satisfy a set of differential-algebraic equations as those appearing in
mechanical systems with constraints. In this section we are going to show that
such equations can be seen as the equations of motion of a constrained Lagrangian system,
on a finite-dimensional manifold, as those defined in \cite{CG07}: the second order constrained systems (SOCS). We shall do that within a variational formalism. 

Also, we shall show that the manifold on which each SOCS is defined is a product of two manifolds: one of them related to the underlying mechanical system, and the other to the thermodynamic counterpart. The latter, as stood out in Remarks \ref{ls1} and \ref{ls2}, is a contact manifold.\newline

\subsection{Second order constrained systems (SOCS)}

\label{secc:C4PG}

In \cite{CG07}, a class of Lagrangian systems with higher order
constraints has been studied. Let us recall the definition of such systems in
the second order case.\newline

Fix a smooth $n$-manifold $Q$ and a function $L:TQ\rightarrow\mathbb{R}$. As
usual, we shall say that $\left(  Q,L\right)  $ is a Lagrangian system and $L$
is its Lagrangian function. Given a curve $\gamma:\left[  t_{1},t_{2}\right]
\rightarrow Q$, recall that an infinitesimal variation of $\gamma$ with fixed
end points, or simply a \emph{variation of }$\gamma$, is a curve $\delta
\gamma:\left[  t_{1},t_{2}\right]  \rightarrow TQ$ such that $\delta
\gamma\left(  t\right)  \in T_{\gamma\left(  t\right)  }Q$, for all
$t\in\left[  t_{1},t_{2}\right]  $, and $\delta\gamma\left(  t_{1,2}\right)  $
belongs to the null distribution of $Q$. In particular, $\gamma=\tau
\circ\delta\gamma$, being $\tau:TQ\rightarrow Q$ the canonical tangent bundle
projection. By $\gamma^{\prime}:\left[  t_{1},t_{2}\right]  \rightarrow TQ$
and $\gamma^{\left(  2\right)  }:\left[  t_{1},t_{2}\right]  \rightarrow
T^{\left(  2\right)  }Q$ we shall denote the velocity and the acceleration of
$\gamma$, respectively, and by $\delta\gamma^{\prime}:\left[  t_{1}%
,t_{2}\right]  \rightarrow TTQ$ the velocity of $\delta\gamma$. Here
$T^{\left(  2\right)  }Q$ denotes the second order tangent bundle of $Q$ (see
\cite{CSC86}).

\begin{defn}
\label{dsocs}Let us consider the triples $\left(  L,C_{K},C_{V}\right)  $ with%
\[
C_{K}\subset T^{\left(  2\right)  }Q\ \ \ \text{and}\ \ C_{V}\subset
TQ\times_{Q}TQ,
\]
where $C_{K}$ is a submanifold and $C_{V}$ is such that, for every $v\in TQ$,
the intersection
\begin{equation}
C_{V}\left(  v\right)  \equiv C_{V}\cap\left(  \left\{  v\right\}  \times
T_{\tau\left(  v\right)  }Q\right)  , \label{C(v)}%
\end{equation}
naturally identified with a subset of $T_{\tau\left(  v\right)  }Q$, is empty
or a linear subspace. We shall refer to these triples as \textbf{second order
constrained systems} \textbf{(SOCS)}, with Lagrangian function $L$,
\textbf{kinematic constraints }$C_{K}$ and \textbf{variational constraints
}$C_{V}$. We shall say that $\gamma:\left[  t_{1},t_{2}\right]  \rightarrow Q$
is a \textbf{trajectory} of $\left(  L,C_{K},C_{V}\right)  $ if the following
conditions are satisfied

\begin{enumerate}
\item $\gamma^{\left(  2\right)  }\left(  t\right)  \in C_{K}$, $\forall
t\in\left(  t_{1},t_{2}\right)  $;

\item for all variations $\delta\gamma$ such that $\left(  \gamma^{\prime
}\left(  t\right)  ,\delta\gamma\left(  t\right)  \right)  \in C_{V}$,
\[
\int_{t_{1}}^{t_{2}}\left\langle dL\left(  \gamma^{\prime}\left(  t\right)
\right)  ,\kappa\left(  \delta\gamma^{\prime}\left(  t\right)  \right)
\right\rangle \ dt=0,
\]
where $\kappa:TTQ\rightarrow TTQ$ is the canonical involution (see \cite{TU99}).
\end{enumerate}
\end{defn}

Let us describe such triples and their related equations in local terms.
Consider a local chart $\left(  U,\varphi\right)  $ of $Q$, with
$\varphi:U\rightarrow\mathbb{R}^{n}$. Given $q\in U$ and $v\in T_{q}U$, write
$\varphi\left(  q\right)  =\left(  q^{1},...,q^{n}\right)  $ and%
\begin{equation}
\varphi_{\ast}\left(  v\right)  =\left(  q^{1},...,q^{n},\dot{q}^{1}%
,...,\dot{q}^{n}\right)  \ \ \ \text{or\ \ \ }\varphi_{\ast}\left(  v\right)
=\left(  q^{1},...,q^{n},\delta q^{1},...,\delta q^{n}\right)  ,
\label{chartPhi}%
\end{equation}
where $\varphi_{\ast}:TU\rightarrow\mathbb{R}^{n}\times\mathbb{R}^{n}$ is the
differential of $\varphi$ (under usual identifications). Sometimes, we shall
also use $q$ and $\left(  q,\dot{q}\right)  $ or $\left(  q,\delta q\right)  $
in order to denote $\varphi\left(  q\right)  $ and $\varphi_{\ast}\left(
v\right)  $, respectively, just for brevity. In this notation, for the local
representative of $\tau$ we have that
\[
\tau\left(  q,\dot{q}\right)  =\tau\left(  q,\delta q\right)  =q.
\]
For a curve $\gamma:\left[  t_{1},t_{2}\right]  \rightarrow Q$, write%
\[%
\begin{array}
[c]{l}%
\varphi\left(  \gamma\left(  t\right)  \right)  =\left(  q^{1}\left(
t\right)  ,...,q^{n}\left(  t\right)  \right)  =q\left(  t\right)  ,\\
\\
\varphi_{\ast}\left(  \gamma^{\prime}\left(  t\right)  \right)  =\left(
q^{1}\left(  t\right)  ,...,q^{n}\left(  t\right)  ,\dot{q}^{1}\left(
t\right)  ,...,\dot{q}^{n}\left(  t\right)  \right)  =\left(  q\left(
t\right)  ,\dot{q}\left(  t\right)  \right)  ,\\
\\
\varphi_{\ast}\left(  \delta\gamma\left(  t\right)  \right)  =\left(
q^{1}\left(  t\right)  ,...,q^{n}\left(  t\right)  ,\delta q^{1}\left(
t\right)  ,...,\delta q^{n}\left(  t\right)  \right)  =\left(  q\left(
t\right)  ,\delta q\left(  t\right)  \right)  ,
\end{array}
\]
in the open set where $\varphi\circ\gamma$ is defined. Finally, for a point
$\eta\in T_{q}^{\left(  2\right)  }Q$, write%
\[
\varphi^{\left(  2\right)  }\left(  \eta\right)  =\left(  q^{1},...,q^{n}%
,\dot{q}^{1},...,\dot{q}^{n},\ddot{q}^{1},...,\ddot{q}^{n}\right)  =\left(
q,\dot{q},\ddot{q}\right)  ,
\]
and for a curve $\gamma$,%
\[
\varphi^{\left(  2\right)  }\left(  \gamma^{\left(  2\right)  }\left(
t\right)  \right)  =\left(  q^{1}\left(  t\right)  ,...,q^{n}\left(  t\right)
,\dot{q}^{1}\left(  t\right)  ,...,\dot{q}^{n}\left(  t\right)  ,\ddot{q}%
^{1}\left(  t\right)  ,...,\ddot{q}^{n}\left(  t\right)  \right)  =\left(
q\left(  t\right)  ,\dot{q}\left(  t\right)  ,\ddot{q}\left(  t\right)
\right)  ,
\]
where $\varphi^{\left(  2\right)  }:T^{\left(  2\right)  }U\rightarrow
\mathbb{R}^{n}\times\mathbb{R}^{n}\times\mathbb{R}^{n}$ is the $2$-lift of
$\varphi$ (again, under usual identifications). In these terms, given a triple
$\left(  L,C_{K},C_{V}\right)  $, if $C_{K}$ is a regular submanifold and
$C_{V}$ is such that the subspaces $C_{V}\left(  v\right)  $ depends smoothly
on $v$, then such subsets are locally given by equations of the form%
\[
w^{a}\left(  q,\dot{q},\ddot{q}\right)  =0\ \ \ \text{and\ \ \ }v_{i}%
^{b}\left(  q,\dot{q}\right)  \,\delta q^{i}=0,
\]
respectively, for certain functions $w^{a}$'s and $v_{i}^{b}$'s (sum over
repeated indices convention is assumed form now on). As a consequence, a curve
$\gamma$ is a trajectory of the triple if and only if

\begin{enumerate}
\item $w^{a}\left(  q\left(  t\right)  ,\dot{q}\left(  t\right)  ,\ddot
{q}\left(  t\right)  \right)  =0$,

\item and, as it is easy to show,%
\begin{equation}
\left(  \frac{d}{dt}\left(  \frac{\partial\left(  L\circ\varphi_{\ast}%
^{-1}\right)  }{\partial\dot{q}^{i}}\left(  q\left(  t\right)  ,\dot{q}\left(
t\right)  \right)  \right)  -\frac{\partial\left(  L\circ\varphi_{\ast}%
^{-1}\right)  }{\partial q^{i}}\left(  q\left(  t\right)  ,\dot{q}\left(
t\right)  \right)  \right)  \delta q^{i}\left(  t\right)  =0 \label{glde}%
\end{equation}
for all functions $\delta q^{i}$ such that%
\[
v_{i}^{b}\left(  q\left(  t\right)  ,\dot{q}\left(  t\right)  \right)
\,\delta q^{i}\left(  t\right)  =0.
\]
Eq. (\ref{glde}) is called \emph{generalized Lagrange-D'Alembert equation}.
\end{enumerate}

\begin{defn}
Given $v\in TQ$ such that $C_{V}\left(  v\right)  $ is not empty, define
\[
F_{V}\left(  v\right)  =\left(  C_{V}\left(  v\right)  \right)  ^{\circ
}\subset T_{\tau\left(  v\right)  }^{\ast}Q.
\]
The union of the subsets $\left\{  v\right\}  \times F_{V}\left(  v\right)  $
defines a subset $F_{V}\subset TQ\times_{Q}T^{\ast}Q$, that we will call the
space of \textbf{constraint forces}.
\end{defn}

\begin{rem}
Since $C_{V}$ and $F_{V}$ are related by annihilation, both of them contain
the same information. So, we can also describe the SOCSs as triples $\left(
L,C_{K},F_{V}\right)  $.
\end{rem}

Particular examples of SOCSs are the \emph{holonomic }and the
\emph{nonholonomic systems} (see \cite{CG07,G09}).
Indeed, consider a nonholonomic system defined by a Lagrangian function $L$
and a set of constraints given by a distribution $\mathcal{D}\subset TQ$.
Define $C_{K}:=(\tau^{(1,2)})^{-1}(\mathcal{D})$ and $C_{V}:=TQ\times
_{Q}\mathcal{D}$, where $\tau^{(1,2)}:T^{(2)}Q\rightarrow TQ$ is the canonical
projection (in coordinates, $\tau^{\left(  1,2\right)  }\left(  q,\dot
{q},\ddot{q}\right)  =\left(  q,\dot{q}\right)  $). Then $(L,C_{K},C_{V})$ is
a SOCS whose trajectories, with initial conditions inside $\mathcal{D}$,
coincide with those of the given nonholonomic system. In the case of a
holonomic system with constraints given by a submanifold $Q_{1}\subset Q$,
define%
\begin{equation}
C_{K}:=(\tau^{(1,2)})^{-1}(TQ_{1})\ \ \ \text{and}\ \ \ C_{V}:=TQ_{1}%
\times_{Q_{1}}TQ_{1}. \label{ckcv}%
\end{equation}
Again, the related SOCS has the same trajectories as the mentioned holonomic
system (for initial conditions inside $TQ_{1}$), i.e. the same trajectories as
the Lagrangian system $\left(  Q_{1},\left.  L\right\vert _{TQ_{1}}\right)  $.
Following similar ideas, \emph{generalized nonholonomic systems }(GNHS) (see
\cite{CG06} and \cite{CIdLdD04}) can also be seen as SOCSs.

\begin{rem}
Notice that for nonholonomic systems, given $q\in Q$ and $v\in T_{q}Q$, we
have that $C_{V}\left(  v\right)  =\mathcal{D}_{q}$. Consequently,
$F_{V}\left(  v\right)  =\mathcal{D}_{q}^{\circ}$ , that is, the constraint
forces vanish on the allowed velocities, which is the content of the
D'Alembert's Principle.\newline
\end{rem}

In what follows, we shall see that the thermo-mechanical systems presented in
Section \ref{Sec:examples} can be seen as SOCSs.\newline

%\begin{thebibliography}{9999}                                                                                             %
%\bibitem {Cr86}M. Crampin, W. Sarlet \& F. Cantrijn,  \textbf{99} (1986), 565-587.
%\bibitem {cg}H. Cendra, S. Grillo, \textit{Generalized nonholonomic mechanics,
%servomechanisms and related brackets}, J. Math. Phys. \textbf{47} (2006), 022902/29.
%\bibitem {hocs}H. Cendra, S. Grillo, \textit{Lagrangian systems with higher
%order constraints}, J. Math. Phys. \textbf{48} (2007), 052904/35.
%\bibitem {her}H. Cendra, A. Ibort, M. de Le\'{o}n, D. de Diego, \textit{A
%generalization of Chetaev's principle for a class of higher order
%non-holonomic constraints}, J. Math. Phys. \textbf{45} (2004), 2785-2801.
%\bibitem {tul}W.M. Tulczyjew, P. Urbanski, \textit{A Slow and Careful Legendre
%Transformation for Singular Lagrangians}, Acta Physica Polonica B \textbf{30}
%(10) (1999), 2909.
%\end{thebibliography}

\subsection{Wagon with internal friction revisited}

\label{ResL:CF}

Let $Q=\mathbb{R}\times\mathbb{R}^{+}\times\mathbb{R}\times\mathbb{R}$, where
$\mathbb{R}^{+}$ is the set of positive real numbers, denote by $\left(
x,T,S,U\right)  $ the points of $Q$ and define $L:TQ\rightarrow\mathbb{R}$ as%
\begin{equation}
L\left(  x,T,S,U,\dot{x},\dot{T},\dot{S},\dot{U}\right)  =\dfrac{m\dot{x}^{2}%
}{2}-U, \label{lw}%
\end{equation}
with $m$ a positive constant. For the Lagrangian system $\left(  Q,L\right)
$, consider the second order kinematic constraints $C_{K}\subset T^{\left(
2\right)  }Q$ given by the points
\[
\left(  x,T,S,U,\dot{x},\dot{T},\dot{S},\dot{U},\ddot{x},\ddot{T},\ddot
{S},\ddot{U}\right)
\]
such that%
\begin{equation}
U=\nu T,\ \ \ S=S_{0}+\nu\ \mathrm{ln}\left(  \dfrac{T}{T_{0}}\right)
\ \ \ \text{and}\ \ \ m\ \ddot{x}\ \dot{x}+\dot{U}=0, \label{kc}%
\end{equation}
where $\nu$, $T_{0}$ and $S_{0}$ are positive constant. If $m$ and $\nu$ are
the mass and the specific heat of the wagon presented in Section \ref{Ej:CcF},
it is clear that $L$ is the difference between its kinetic and its internal
energy, and the equations in (\ref{kc}) correspond to the state equations of
the thermodynamic counterpart and the conservation of the total energy $E$ of
the wagon [see Eq. (\ref{cec})].

Finally, consider the variational constraints $C_{V}\subset TQ\times_{Q}TQ$
defined by the points
\[
\left(  x,T,S,U,\dot{x},\dot{T},\dot{S},\dot{U},\delta x,\delta T,\delta
S,\delta U\right)
\]
such that
\begin{equation}
\delta U=\nu\delta T,\ \ \ \delta S=\nu\frac{\delta T}{T}\ \ \ \text{and\ }%
\ \ \ \mu\dot{x}\,\delta x=\delta U, \label{vc}%
\end{equation}
with $\mu$ another positive constant. If $\mu$ is the friction coefficient
between the axles and the chassis of the wagon, the last equation say that the
infinitesimal variation of the internal energy $U$ of the wagon is due to the
infinitesimal work made by the friction force.

It can be shown that the trajectories of the triple $\left(  L,C_{K}%
,C_{V}\right)  $ are in bijection with the curves $t\longmapsto\left(
x\left(  t\right)  ,\dot{x}\left(  t\right)  ,T\left(  t\right)  ,S\left(
t\right)  ,U\left(  t\right)  \right)  $ found in Section \ref{Ej:CcF}. In
fact, the generalized Lagrange-D'Alembert Equation (\ref{glde}) for $\left(
L,C_{K},C_{V}\right)  $ reduces to
\begin{equation}
-m\ddot{x}\,\delta x-\delta U=0.\label{Ec:CF1 var}%
\end{equation}
Using the latter and the variational constraint $\mu\dot{x}\,\delta x=\delta
U$ [see Eq. (\ref{vc})], the Newton's equation for the wagon $m\ddot{x}%
=-\mu\dot{x}$ follows [recall Eq. (\ref{AH:Newq})]; and using the kinematic
constraints given by Eq. (\ref{kc}), we have the Eqs. (\ref{Ec:CF EE}) and
(\ref{cec}). In conclusion, we can describe the thermo-mechanical system
presented in Section \ref{Ej:CcF} as the SOCS $\left(  L,C_{K},C_{V}\right)  $
given by Eqs. (\ref{lw}), (\ref{kc}) and (\ref{vc}). A similar assertion can
be made about the thermo-mechanical system of Section \ref{wter}, provided we
assume the Fourier law holds. We just must replace the condition $m\ \ddot
{x}\ \dot{x}+\dot{U}=0$ by [recall Eq. (\ref{nce})]
\[
m\ \ddot{x}\ \dot{x}+\dot{U}=\kappa\,\left(  T_{b}-T\right)  .
\]
In both cases, the Second Law is automatically satisfied. 

On the other hand, if no further information is given for the heat exchange, then we do have a
SOCS again, but its equations of motion do not determine completely the
trajectories. In addition, in such a case, we must ask the trajectories to satisfy
the Second Law.

To end this subsection, note that the manifold $Q$ can be written as a Cartesian product of two manifolds: $\cal{M}=\mathbb{R}$, related to the mechanical degrees of freedom, and $\cal{T}=\mathbb{R}^+ \times \mathbb{R} \times \mathbb{R}$, related to the thermodynamical ones (namely $U$, $T$ and $S$). The latter, as explained in Remark \ref{ls1}, is a $3$-dimensional contact manifold with contact form $\theta=dU-T dS$. Note also that some of the conditions that define $C_K$, the state equations, give rise to a Legendre submanifold of $\cal{T}$. As we shall see below, this is not a peculiarity of the present example, but a common characteristic of all the examples we introduced in this paper.  

\subsection{Vertical piston revisited}

\label{ResL:EA}

\subsubsection{The adiabatic case}

Let $Q=\mathbb{R}\times\mathbb{R}^{+}\times\mathbb{R}^{+}\times\mathbb{R}%
^{+}\times\mathbb{R}\times\mathbb{R}$, denote by $\left(  x,P,T,V,S,U\right)
$ the points of $Q$ and define $L:TQ\rightarrow\mathbb{R}$ as%
\[
L\left(  x,P,T,V,S,U,\dot{x},\dot{P},\dot{T},\dot{V},\dot{S},\dot{U}\right)
=\dfrac{m\dot{x}^{2}}{2}-mgx-U,
\]
with $m$ and $g$ positive constants. Consider the submanifold $C_{K}\subset
T^{\left(  2\right)  }Q$ given by the equations%
\begin{equation}
PV=N_{0}RT,\quad U=\alpha N_{0}RT,\quad Ax=V, \label{q1}%
\end{equation}
\begin{equation}
\label{q2}\ S=S_{0} \quad\text{and} \quad PV^{\gamma}=k ,
\end{equation}
and $C_{V}\subset TQ\times_{Q}TQ$ given by%
\begin{equation}
\label{p1}\delta P \, V+P\, \delta V =N_{0}R\, \delta T,\quad\delta U=\alpha
N_{0}R\, \delta T,\ \ A\,\delta x=\delta V ,
\end{equation}
\[
\delta S =0\quad\text{and }\quad\delta P \, V^{\gamma}+\gamma P\,V^{\gamma
-1}\,\delta V=0\ ,
\]
where $N_{0}$, $R$, $\alpha$, $S_{0}$, $k$ and $A$ are positive constants, and
$\gamma=\frac{\alpha+1}{\alpha}$.

The generalized Lagrange-D'Alembert equation for $\left(  L,C_{K}%
,C_{V}\right)  $ is%
\[
-m\ddot{x}\,\delta x-mg\,\delta x-\delta U=0.
\]
Using the variational constraints, it is easy to show that $\delta
U=-P\,\delta V=-PA\,\delta x$. Then, the equation above reduces to%
\[
-m\ddot{x}-mg+PA=0.
\]
This equation together with the kinematic constraints in Eqs. (\ref{q1}) and
(\ref{q2}) are exactly the Equations (\ref{Ec:GI1}), (\ref{Ec:GI2}),
(\ref{Ec:GIA}), (\ref{Scero}), (\ref{Ec:EA Newton}) and (\ref{Ec:EA VinG})
obtained for the thermo-mechanical system presented in Section \ref{adcase}.
Thus, the adiabatic vertical piston is a SOCS.

\begin{rem}
Note that Eqs. (\ref{q1}) and (\ref{q2}) defines a submanifold $Q_{1}\subset
Q$ in terms of which we can define
\[
C_{K}:=(\tau^{(1,2)})^{-1}(TQ_{1})\ \ \ \text{and}\ \ \ C_{V}\left(  v\right)
:=T_{q}Q_{1},
\]
for all $q\in Q_{1}$ and $v\in T_{q}Q$. Taking into account the discussion
around Eq. (\ref{ckcv}), the equation above says that the SOCS $\left(
L,C_{K},C_{V}\right)  $ can be seen as a holonomic system (with constraints
given by $Q_{1}$), i.e. it is equivalent to the Lagrangian system $\left(
Q_{1},\left.  L\right\vert _{TQ_{1}}\right)  $. Since, in addition, $\left.
L\right\vert _{TQ_{1}}$ is hyperregular, such a HOCS is equivalent to a
Hamiltonian system, as it was established in Section \ref{revq}.
\end{rem}

Recall that condition $Ax=V$ can be replaced by the conservation of the
quantity%
\[
E=\frac{m\dot{x}^{2}}{2}+mgx+U,
\]
[see Section \ref{enco}], which gives the second order constraint%
\[
m\ddot{x}\,\dot{x}+mg\,\dot{x}+\dot{U}=0
\]
[see Eq. (\ref{sicon})]. It can be shown that, if we change $Ax=V$ by the
equation above, and change $A\,\delta x=\delta V$ by%
\[
-PA\,\delta x=\delta U,
\]
the new SOCS has the same trajectories as the previous one. The last equation
says that the infinitesimal variation of the internal energy is due to the
infinitesimal work made by the pressure.

As in the previous example, let us note that the manifold $Q$ can be written as a Cartesian product of two manifolds. In this case, $\cal{M}=\mathbb{R}$ and $\cal{T}=\mathbb{R}^{+}\times\mathbb{R}^{+}\times\mathbb{R}%
^{+}\times\mathbb{R}\times\mathbb{R}$, where the latter is a contact manifold with global Darboux coordinates $(P,-T,V,S,U)$ and contact form $\theta=dU - T dS + P dV$.

\subsubsection{The isothermal case}

Take $Q$ and $L$ as above, and define $C_{K}\subset T^{\left(  2\right)  }Q$
by the equations (\ref{q1}) plus
\begin{equation}
\label{ck}S =S_{0}+N_{0}R\ \mathrm{ln}\left(  \dfrac{T^{\alpha}V}%
{T_{0}^{\alpha}V_{0}}\right)  ,\quad T=T_{0}%
\end{equation}
and [see Eq. (\ref{noncon})]%
\[
m\ddot{x}\,\dot{x}+mg\,\dot{x}+\dot{U}=T\,\dot{S}.
\]
Finally, define $C_{V}\subset TQ\times_{Q}TQ$ by the equations (\ref{p1})
plus
\begin{equation}
\delta S=N_{0}R\left(  \frac{\delta V}{V}-\alpha\frac{\delta T}{T}\right)  ,
\quad\delta T=0 \label{cv}%
\end{equation}
and%
\[
-PA\,\delta x=\delta U.
\]
Again, $N_{0}$, $R$, $\alpha$, $T_{0}$, $S_{0}$, $V_{0}$ and $A$ are positive
constants. Note that all the variations must vanish. So, the generalized
Lagrange-D'Alembert equation for $\left(  L,C_{K},C_{V}\right)  $ is trivial.
The constraint equations are the only relevant equations. They give precisely
(\ref{rei}) and (\ref{nei}). As a consequence, we can describe the
thermo-mechanical system presented in Section \ref{isot} as a SOCS [see Remark \ref{ls2}].

\subsubsection{Another thermodynamic potentials}

An alternative description of the previous thermo-mechanical system can be
given by considering the \emph{Helmholtz free energy} $H:=U-TS$ when defining
the Lagrangian function. In other words, let us consider $Q$ as above, but
define%
\[
L\left(  x,P,T,V,S,U,\dot{x},\dot{P},\dot{T},\dot{V},\dot{S},\dot{U}\right)
:=\dfrac{m\dot{x}^{2}}{2}-mgx-U+TS.
\]
Also, define $C_{K}$ by (\ref{q1}) and (\ref{ck}), and $C_{V}$ by (\ref{p1})
and (\ref{cv}). Now, the generalized Lagrange-D'Alembert equation, together
with the constraints $T=T_{0}$ and $Ax=V$, give exactly the Eq. (\ref{rei}).
The Eqs. (\ref{nei}) are obtained from (\ref{q1}) and (\ref{ck}). We can say
that it is more natural to use $H$ instead of $U$ in order to define the
present thermo-mechanical system as a SOCS, in the sense that, under such a
choice, the variational condition turns out to be non trivial. Nevertheless,
the only thermodynamical potential that we shall consider from now on (to
construct the Lagrangian functions) will be the internal energy.

\subsubsection{Vertical piston with dissipation}

\label{verdis}

Let $Q=\mathbb{R}\times\mathbb{R}^{+}\times\mathbb{R}^{+}\times\mathbb{R}%
^{+}\times\mathbb{R}\times\mathbb{R}\times\mathbb{R}^{+}\times\mathbb{R}%
\times\mathbb{R}$, denote by $\left(  x,P,T,V,S,U;T_{c},S_{c},U_{c}\right)  $
the points of $Q$ and define $L:TQ\rightarrow\mathbb{R}$ as%
\[
L\left(  x,P,T,V,S,U,T_{c},S_{c},U_{c},\dot{x},\dot{P},\dot{T},\dot{V},\dot
{S},\dot{U};\dot{T}_{c},\dot{S}_{c},\dot{U}_{c}\right)  =\dfrac{m\dot{x}^{2}%
}{2}-mgx-U-U_{c},
\]
with $m$ and $g$ positive constants. Consider the submanifold $C_{K}\subset
T^{\left(  2\right)  }Q$ given by the equations (\ref{q1}) plus
\[
S=S_{0}+N_{0}R\mathrm{ln}\left(  \dfrac{T^{\alpha}V}{T_{0}^{\alpha}V_{0}%
}\right)  ,\quad U_{c}=\nu T_{c},\ \quad S_{c}=\nu\mathrm{ln}\left(
\dfrac{T_{c}}{T_{c,0}}\right)  +S_{c,0}%
\]
and [see (\ref{ce}) and (\ref{Fl})]
\[
\dot{E}_{mec}+\dot{U}_{c}=-\kappa\mathcal{A}(T-T_{c})\quad\text{and }\quad
\dot{E}_{mec}+\dot{U}+\dot{U}_{c}=0.
\]
On the other hand, take $C_{V}\subset TQ\times_{Q}TQ$ given by Eqs. (\ref{p1})
plus
\[
\delta S=N_{0}R\left(  \frac{\delta V}{V}+\alpha\frac{\delta T}{T}\right)
,\qquad\delta U_{c}-\nu\ \delta T_{c}=0,\qquad\delta S_{c}=\nu\left(
\frac{\delta T_{c}}{T_{c}}\right)  ,
\]%
\begin{equation}
\delta U=-P\ \delta V\qquad\text{and}\qquad\delta U_{c}=\mu\dot{x}\delta x,
\label{cv2}%
\end{equation}
where $N_{0}$, $R$, $\alpha$, $S_{0}$, $S_{c,0}$, $T_{0}$, $T_{c,0}$, $V_{0}$,
$k$, $A$, $\mu$ and $\nu$ are positive constants, and $\mathcal{A}$ is the
area through which heat flows.

The generalized Lagrange-D'Alembert equation for $\left(  L,C_{K}%
,C_{V}\right)  $ is%
\[
-m\ddot{x}\,\delta x-mg\,\delta x-\delta U-\delta U_{c}=0.
\]
Using the variational constraints in Eqs (\ref{p1}) and (\ref{cv2}) it is easy
to show that the equation above reduces to the Newton's Eq. (\ref{Ec:ECF1}).
So this equation together with the kinematic constraints defined above are
exactly the Equations involved in the thermo-mechanical system presented in
Section \ref{Sec:EVDF}, and then the vertical piston with dissipation is also
a SOCS.

As for the wagon in a thermal bath, if we do not assume the Fourier's law, we
have again a SOCS for which, in addition, the Second Law must be requested (because
it is not automatically satisfied).

Completing this description, let us note again that the manifold $Q$ can be written as a Cartesian product of $\cal{M}=\mathbb{R}$ and $\cal{T}=\mathbb{R}^{+} \times \mathbb{R}^{+} \times \mathbb{R}^{+} \times \mathbb{R} \times \mathbb{R} \times\mathbb{R}^{+} \times \mathbb{R} \times \mathbb{R}$. Nonetheless, the latter can also be decomposed as a product of two contact manifolds $\mathcal{T}_{1}$ and $\mathcal{T}_{2}$, i.e. this is a composite system [see Remark \ref{comp}], where $\mathcal{T}_{1}=\mathbb{R}^{+}\times\mathbb{R}^{+}\times\mathbb{R}^{+}\times\mathbb{R}\times\mathbb{R}$ represents the ideal gas and ${\mathcal{T}}_{2}=\mathbb{R}^{+}\times\mathbb{R}\times\mathbb{R}$ represents the container, and the contact form is given by $\theta=dU - T dS + P dV + dU_c -T_c d S_c$ [see Remark \ref{comp2}].
%Then $(x,P,T,V,S,U)$ and $(T_{c},S_{c},U_{c})$ are the global Darboux coordinates of ${\mathcal{T}}_{1}$ and ${\mathcal{T}}_{2}$ respectively
\bigskip

\subsection{Variational formulation of thermo-mechanical systems}

Taking into account the previous examples (and our knowledge of mechanical and
thermodynamic systems -see Appendix-), we shall develop a proposal to describe
the thermo-mechanical systems as SOCSs. In order to do that, let us summarize
what the above SOCSs $\left(  L,C_{K},C_{V}\right)  $ have in common.

\subsubsection{The configuration space}

The manifold $Q$ can be written as a Cartesian product $Q=\mathcal{M}%
\times\mathcal{T}$, where $\mathcal{M}$ and $\mathcal{T}$ are manifolds that
define the mechanical and the thermodynamic observables, respectively. Also,
$\mathcal{T}$ is an open subset of $\mathbb{R}^{2d+1}$, for some
$d\in\mathbb{N}$, and, consequently, $\mathcal{T}$ is a contact manifold.
Moreover, $\mathcal{T}$ has a distinguished set of global coordinates. As in
the Appendix, let us denote the latter by
\begin{equation}
\left(  x_{1},...,x_{d-1},T,y_{1},...,y_{d-1},S,U\right)  . \label{gloco}%
\end{equation}

\subsubsection{The Lagrangian function}

The Lagrangian $L:TQ\rightarrow\mathbb{R}$, identifying $T\left(
\mathcal{M}\times\mathcal{T}\right)  $ with $T\mathcal{M}\times T\mathcal{T}$,
can be written as a sum
\[
L\left(  a,b\right)  =L_{mec}\left(  a\right)  -U\left(  \tau\left(  b\right)
\right)  ,\ \ \ \forall a\in T\mathcal{M},\ \ b\in T\mathcal{T},
\]
for some function $L_{mec}:T\mathcal{M}\rightarrow\mathbb{R}$. Here,
$\tau:T\mathcal{T}\rightarrow\mathcal{T}$ is the canonical projection. For
instance, for the vertical piston of Section \ref{Ej:EVA}, we have
$\mathcal{M}=\mathbb{R}$ and $L_{mec}:T\mathbb{R\rightarrow R}$ such that
\[
L_{mec}\left(  x,\dot{x}\right)  =\frac{m\dot{x}^{2}}{2}-mgx.
\]

\begin{rem}
Note that the energy\footnote{Recall that, given a Lagrangian function
$L:TQ\rightarrow\mathbb{R}$, its \emph{energy }$E:TQ\rightarrow\mathbb{R}$ is
given by%
\[
E\left(  v\right)  =\left\langle \mathbb{F}L\left(  v\right)  ,v\right\rangle
-L\left(  v\right)  ,\ \ \ \forall v\in TQ,
\]
being $\mathbb{F}L:TQ\rightarrow T^{\ast}Q$ the fiber derivative of $L$, i.e.
the Legendre transformation related to $L$.} $E:T\mathcal{M}\times
T\mathcal{T}\rightarrow\mathbb{R}$ of $L$ is%
\[
E\left(  a,b\right)  =E_{mec}\left(  a\right)  +U\left(  \tau\left(  b\right)
\right)  ,
\]
being $E_{mec}:T\mathcal{M}\rightarrow\mathbb{R}$ the energy of $L_{mec}$. For
instance, for the vertical piston,
\[
E_{mec}\left(  x,\dot{x}\right)  =\frac{m\dot{x}^{2}}{2}+mgx.
\]

\end{rem}

\subsubsection{Kinematic constraints}

Identifying $T^{\left(  2\right)  }Q$ and $T^{\left(  2\right)  }%
\mathcal{M}\times T^{\left(  2\right)  }\mathcal{T}$, the kinematic
constraints always satisfy%
\[
C_{K}\subset\left(  T^{\left(  2\right)  }\mathcal{M}\times T^{\left(
2\right)  }\mathcal{N}\right)  \cap C_{E, \dbar
\mathbb{Q}},
\]
where $\mathcal{N}\subset\mathcal{T}$ is a Legendre submanifold related to the
(thermodynamic) state equations (see Remarks \ref{ls1} and \ref{ls2}) and
$C_{E,\dbar \mathbb{Q}}$ is defined as follows.

Consider the canonical projections $\tau^{\left(  1,2\right)  }:T^{\left(
2\right)  }Q\rightarrow TQ$ and $\tau^{\left(  2\right)  }:T^{\left(
2\right)  }Q\rightarrow Q$, and the canonical immersion\footnote{The local
representative of $j^{\left(  2\right)  }$ (in the above mentioned local
charts) is
\[
j^{\left(  2\right)  }\left(  q,\dot{q},\ddot{q}\right)  =\left(  q,\dot
{q},\dot{q},\ddot{q}\right)  .
\]
} $j^{\left(  2\right)  }:T^{\left(  2\right)  }Q\rightarrow TTQ$. Finally,
define $C_{E,\dbar \mathbb{Q}}$ as the
submanifold given by the points $\eta\in T^{\left(  2\right)  }Q$ such that
\begin{equation}
\left\langle dE\left(  \tau^{\left(  1,2\right)  }\left(  \eta\right)
\right)  ,j^{\left(  2\right)  }\left(  \eta\right)  \right\rangle
=\left\langle  \dbar \mathbb{Q}\left(
\tau^{\left(  2\right)  }\left(  \eta\right)  \right)  ,\tau^{\left(
1,2\right)  }\left(  \eta\right)  \right\rangle , \label{cke}%
\end{equation}
being $E$ the energy of $L$ and $\dbar
\mathbb{Q}$ a $1$-form on $Q$. It would be enough to take $\dbar \mathbb{Q}$ as a $1$-form on $\mathcal{N}$. We take it this
way just for simplicity.

Eq. (\ref{cke}) is a second order constraint corresponding to the energy
conservation assumption \textbf{ECP} (see Section \ref{enco}), where
$\dbar \mathbb{Q}$ represents the infinitesimal
heat exchange between the system and the environment. For instance, for the
wagon with internal friction in a thermal bath (see Section \ref{wter}),
\[
C_{E, \dbar \mathbb{Q}}=\left\{  \left(
x,T,S,U,\dot{x},\dot{T},\dot{S},\dot{U},\ddot{x},\ddot{T},\ddot{S},\ddot
{U}\right)  :m\ \ddot{x}\ \dot{x}+\dot{U}=\kappa\,\left(  T_{b}-T\right)
\right\}  .
\]

Depending on the system, beside those defined by $\mathcal{N}$ and
$C_{E, \dbar \mathbb{Q}}$, additional constraints
use to be present.

For instance, for all the versions of the vertical piston we have the
geometrical constraint $Ax=V$, which defines a submanifold $C\subset
T^{\left(  2\right)  }\mathcal{M}\times T^{\left(  2\right)  }\mathcal{T}$.
For the adiabatic version we have the constraint $PV^{\gamma}=k$, and for the
version with dissipation (see Section \ref{Sec:EVDF}) we have%
\[
\dot{T}=\frac{\kappa\mathcal{A}}{\alpha N_{0}R}\,\left(  T_{c}-T\right)  ,
\]
corresponding to the Fourier's law. The last two constraints only involve
thermodynamic observables, and define submanifolds $C_{K}^{ter}\subset
T^{\left(  2\right)  }\mathcal{N}$. In the examples given in this paper there
are no constraints on the mechanical observables alone. Constraints of this
kind would define a submanifold $C_{K}^{mec}\subset T^{\left(  2\right)
}\mathcal{M}$. Thus, in general, we can say that $C_{K}$ is of the form
\[
C_{K}=C\cap\left(  C_{K}^{mec}\times C_{K}^{ter}\right)  \cap
C_{E, \dbar \mathbb{Q}}.
\]

\subsubsection{Variational constraints}

On one hand, related to the thermodynamic observables only, each state
equation gives rise, by derivation, to a variational constraint. For instance,
the ideal gas equation $PV=N_{0}RT$ gives rise to the variational constraint%
\[
\delta P\,V+P\,\delta V=N_{0}R\,\delta T.
\]
This is why for each $q=\left(  m,n\right)  \in\mathcal{M}\times\mathcal{T}$
and $v\in T_{q}Q=T_{m}\mathcal{M}\times T_{n}\mathcal{T}$, we have that
\[
C_{V}\left(  v\right)  \subset T_{m}\mathcal{M}\times T_{n}\mathcal{N}%
\]
On the other hand, consider a fiber-preserving map $\mathcal{F}:TQ\rightarrow
T^{\ast}Q$ such that
\begin{equation}
\left\langle \mathcal{F}\left(  v\right)  ,\left(  a,b\right)  \right\rangle
:=\left\langle i_{\tau\left(  b\right)  }^{\ast}\mathcal{F}\left(  v\right)
,a\right\rangle +\delta U,\ \ \ \forall\left(  a,b\right)  \in T\mathcal{M}%
\times T\mathcal{T}, \label{fvw}%
\end{equation}
where%
\[
b=\left(  \delta x_{1},...,\delta x_{d-1},\delta T,\delta y_{1},...,\delta
y_{d-1},\delta S,\delta U\right)  \in T\mathcal{T},
\]
and $i_{n}:\mathcal{M}\rightarrow Q:m\mapsto\left(  m,n\right)  $. Now,
define
\begin{equation}
C_{\mathcal{F}}:=\left\{  \left(  v,w\right)  \in TQ\times_{Q}TQ:\left\langle
\mathcal{F}\left(  v\right)  ,w\right\rangle =0\right\}  , \label{cvf}%
\end{equation}
and [recall Eq. (\ref{C(v)})]%
\begin{equation}
C_{\mathcal{F}}\left(  v\right)  =C_{\mathcal{F}}\cap\left(  \left\{
v\right\}  \times T_{\tau\left(  v\right)  }Q\right)  , \label{cvfv}%
\end{equation}
for each $v\in TQ$. For the wagon with internal friction, the variational
constraint $-\mu\dot{x}\,\delta x+\delta U=0$ can be described by the subset
$C_{\mathcal{F}}$ with (under usual identifications)
\begin{equation}
\mathcal{F}:\left(  x,T,S,U,\dot{x},\dot{T},\dot{S},\dot{U}\right)
\longmapsto\left(  x,T,S,U,-\mu\dot{x},0,0,1\right)  . \label{fwif}%
\end{equation}
For the adiabatic vertical piston, the subset $C_{\mathcal{F}}$ with
\[
\mathcal{F}:\left(  x,P,T,V,S,U,\dot{x},\dot{P},\dot{T},\dot{V},\dot{S}%
,\dot{U}\right)  \longmapsto\left(  x,P,T,V,S,U,PA,0,0,0,0,1\right)
\]
describes the variational constraint\ $PA\,\delta x+\delta U=0$. Note that
$\mathcal{F}$ describe the external force acting on the underlying mechanical
system. It can be shown that
\[
C_{V}\left(  v\right)  \subset\left(  T_{m}\mathcal{M}\times T_{n}%
\mathcal{N}\right)  \cap C_{\mathcal{F}}\left(  v\right)
\]
in all of the above examples, for some function $\mathcal{F}$. Of course,
additional variational constraints are usually present, but they depend on each
particular system.

\subsubsection{The Second Law condition}

As we said at the end of Section \ref{verdis}, among all the trajectories
\[
\gamma:t\mapsto\gamma\left(  t\right)  =\left(  m\left(  t\right)  ,n\left(
t\right)  \right)  \in\mathcal{M}\times\mathcal{N}\subset Q,
\]
we have to choose those for which the Second Law holds
(unless such a law is automatically satisfied). This would mean to impose the
additional condition [see Eqs. (\ref{sl1}) and (\ref{sl2})]
\begin{equation}
\left\langle dS\left(  n\left(  t\right)  \right)  ,\dot{n}\left(  t\right)
\right\rangle \geq\left\langle \frac{ \dbar\mathbb{Q}\left(  \gamma\left(  t\right)  \right)  }{T\left(
n\left(  t\right)  \right)  },\left(  0,\dot{n}\left(  t\right)  \right)
\right\rangle . \label{cond}%
\end{equation}

\subsubsection{Thermo-mechanical systems as SOCSs}

Now, we can give a definition of a thermo-mechanical system in terms of SOCSs.

\begin{defn}
We shall say that a SOCS $\left(  L,C_{K},C_{V}\right)  $ on $Q$ is a
\textbf{thermo-mechanical system} if:

\begin{itemize}
\item There exist manifolds $\mathcal{M}$ and $\mathcal{T}$ such that
$Q=\mathcal{M}\times\mathcal{T}$, being $\mathcal{T}$ an open submanifold of
$\mathbb{R}^{2d+1}$, for some $d\in\mathbb{N}$, with distinguished global
coordinates. We shall denote the latter as in Eq. (\ref{gloco}).

\item There exists a function $L_{mec}:T\mathcal{M}\rightarrow\mathbb{R}$ such
that, using the projections $p^{\mathcal{M},\mathcal{T}}:Q\rightarrow
\mathcal{M},\mathcal{T}$ and $\tau:T {\mathcal{T}}\to{\mathcal{T}}$,%
\[
L=L_{mec}\circ p_{\ast}^{\mathcal{M}}-U\circ\tau\circ p_{\ast
}^{\mathcal{T}}.
\]

\item There exists a Legendre submanifold $\mathcal{N}\subset\mathcal{T}$,
w.r.t. the contact form%
\[
\theta:=dU-T\ dS+%
%TCIMACRO{\dsum \nolimits_{i=1}^{d-1}}%
%BeginExpansion
{\displaystyle\sum\nolimits_{i=1}^{d-1}}
%EndExpansion
x_{i}\,dy_{i},
\]
and a $1$-form $\dbar \mathbb{Q}\in\Omega
^{1}\left(  Q\right)  $ such that%
\[
C_{K}\subset\left(  T^{\left(  2\right)  }\mathcal{M}\times T^{\left(
2\right)  }\mathcal{N}\right)  \cap C_{E, \dbar
\mathbb{Q}},
\]
where $E$ is the energy of $L$ and $C_{E, \dbar \mathbb{Q}}$ is given by Eq. (\ref{cke}).

\item For each $q=\left(  m,n\right)  \in\mathcal{M}\times\mathcal{T}$ and
$v\in T_{m}\mathcal{M}\times T_{n}\mathcal{T}$,%
\[
C_{V}\left(  v\right)  \subset\left(  T_{m}\mathcal{M}\times T_{n}%
\mathcal{N}\right)  \cap C_{\mathcal{F}}\left(  v\right)  ,
\]
where $C_{\mathcal{F}}\left(  v\right)  $ is given by Eqs. (\ref{fvw}),
(\ref{cvf}), and (\ref{cvfv}).

The trajectories of $\left(  L,C_{K},C_{V}\right)  $ are those of Definition
\ref{dsocs} that also fulfill the condition (\ref{cond}).
\end{itemize}
\end{defn}

\bigskip

As an example, we shall consider the dissipative vertical piston as already presented, but now immersed in a thermal bath of constant temperature $T_{b}$.

\begin{itemize}
\item Let $\mathcal{M}=\mathbb{R}^{+}$ and $\mathcal{T}=\mathbb{R}^{+}%
\times\mathbb{R}^{+}\times\mathbb{R}^{+}\times\mathbb{R}\times\mathbb{R}%
\times\mathbb{R}^{+}\times\mathbb{R}\times\mathbb{R}$
%\times  \mathbb{R}^+ \times \mathbb{R} \times \mathbb{R}$
. Denote
\[
(x,P,T,V,S,U;T_{c},S_{c},U_{c})
\]
%; T_b , S_b , U_b)$
to the points in $Q=\mathcal{M}\times\mathcal{T}$. As already shown at the end of Subsection \ref{verdis}, ${\mathcal{T}}$ can be decomposed as a product of two manifolds
(${\mathcal{T}}={\mathcal{T}}_{1}\times{\mathcal{T}}_{2})$, where $(x,P,T,V,S,U)$ and $(T_{c},S_{c},U_{c})$
are the global Darboux coordinates of ${\mathcal{T}}_{1}$ and ${\mathcal{T}%
}_{2}$,
%and ${\mathcal T}_3$
respectively.

\item Take $L_{mec}=\dfrac{m\dot{x}^{2}}{2}-mgx$, $U_{tot}=U+U_{c}$ and define
\[
L=L_{mec}-U_{tot}
\]

\item Labeling $\theta$ and $\theta_{c}$ to the contact forms of
${\mathcal{T}}_{1}$ and ${\mathcal{T}}_{2}$ respectively, the contact form of
${\mathcal{T}}$, according to Eq. (\ref{tita2}), is given by $\theta
_{tot}=\theta+\theta_{c}$. The Legendre submanifold $\mathcal{N}%
\subset\mathcal{T}$ w.r.t $\theta_{tot}$ is given by the state equations of
the ideal gas
\[
PV=N_{0}RT,\qquad U=\alpha N_{0}RT,\qquad S=S_{0}+N_{0}R\ \mathrm{ln}\left(
\dfrac{T^{\alpha}V}{T_{0}^{\alpha}V_{0}}\right)  ,
\]
and the state equations of the container
\[
U_{c}=\nu_{c}T_{c},\qquad S_{c}=\nu_{c}\mathrm{ln}\left(  \dfrac{T_{c}%
}{T_{c,0}}\right)  +S_{c,0}.
\]
%and the state equations of the bath
%$$ U_b=\nu_b T_b,\qquad S_b=\nu_b \mathrm{ln}\left( \dfrac{T_b }{T_{b,0}}\right)+S_{b,0}.$$
The constraint $C_{E, \dbar\mathbb{Q}}$ associated
with the \textbf{ECP} is resumed in the equation
\[
m\ \ddot{x}\ \dot{x}+m\ g\ \dot{x}+\dot{U}+\dot{U_{c}}=\kappa_{e}%
\mathcal{A}_{e}\,\left(  T_{b}-T_{c}\right)  ,
\]
where $\kappa_{e}$ is the conduction coefficient between the
container and the bath, and $\mathcal{A}_{e}$ is the area though which heat
flows from the container to the bath. Both are being considered as constants.

The additional constraints that completes the description of $C_{K}$ are given
by the geometrical configuration and the Fourier's law:
\[
Ax=V \qquad\text{ and } \qquad\dot{U}=-\kappa_{i} \mathcal{A}_{i}(T-T_{c}),
\]
where $\kappa_{i}$ is the constant conduction coefficient between the gas and
the contained. The term $\mathcal{A}_{i}$ is the area though which heat flows from
the gas to the container, which depends linearly on $x$.

\item If we take the same variational constraints as in Section \ref{verdis},
$C_{\mathcal{F}}$ (under usual identifications) is given by
$$
\begin{tikzcd}
\mathcal{F}:\left(  x,P,T,V,S,U; T_c , S_c , U_c,\dot{x},\dot{P},\dot
{T},\dot{V},\dot{S},\dot{U}; \dot{T_c},\dot{S_c},\dot{U_c}\right) \arrow{d} \\
  \left(  x,P,T,V,S,U; T_c , S_c , U_c,-\mu\dot{x}+PA,0,0,0,0,1;0,0,1\right) 
\end{tikzcd}
$$

%\begin{equation*}
%\mathcal{F}:\left(  x,P,T,V,S,U; T_c , S_c , U_c,\dot{x},\dot{P},\dot
%{T},\dot{V},\dot{S},\dot{U}; \dot{T_c},\dot{S_c},\dot{U_c}\right)  \longmapsto\left(  x,P%
%,T,V,S,U; T_c , S_c , U_c,-\mu\dot{x}+PA,0,0,0,0,1,0,0,1\right)
%\end{equation*}

Finally, the trajectories that do not satisfy the inequality (\ref{cond}) must
be discarded.\newline
\end{itemize}

\section{Conlusions and future work}

In this paper, we have studied a class of physical systems that combine (a finite number of) mechanical and thermodynamical degrees of freedom: the thermo-mechanical systems. We have taken a special care in deducing the evolution equations of the involved observables, for which we only used the Newton's Law and the First Law of Thermodynamics. These evolution equations have been studied in detail in several examples. Also, observing that such equations are similar to the equations of motion of a constrained mechanical system, we proposed a description of the thermo-mechanical systems in terms of Lagrangian systems with second order constraints: the SOCSs. Moreover, we characterized the manifolds in which such SOCS are defined as Cartesian products of two manifold: one of them is related to the mechanical degrees of freedom, and the other to the thermodynamical ones. The latter, in turn, is a contact manifold $\cal{T}$ with a distinguished contact form. Let us also mention that the kinematical constraints, related to the state equations of the thermodynamical counterpart, define a Legendre submanifold of $\cal{T}$.

In a forthcoming paper, we shall give another description of the thermo-mechanical systems, combining the  formulation of thermodynamics present in \cite{Mr91,Mr001}, in terms of Legendre submanifolds, and formulation of mechanics shown in \cite{TS72,Tu77,TU99}, in terms of Lagrangian submanifolds.

\section*{Acknowledgments}

The authors thank CONICET for its financial support.
%\appendix

\begin{appendices}
\setcounter{secnumdepth}{0}
\section[: Brief background on thermodynamics]{Appendix: Brief background on thermodynamics}

Below, we introduce the basic notation and terminology on thermodynamics that
we shall use along all of the paper, and recall some fundamental concepts on
the subject (see \cite{BBMS,Fermi,Planck,Callen}).

\bigskip

$\ast$ A thermodynamic system is typically defined by $2d+1$ observables,
which we shall denote $x_{1},...,x_{d-1},T,y_{1},...,y_{d-1},S$ and $U$. The
$x_{i}$'s and $T$ (resp. $y_{i}$'s, $S$ and $U$) are called \emph{intensive}
(resp. \emph{extensive}) \emph{variables}. The extensive variables depend on
the \textquotedblleft size\textquotedblright\ of the system, while the
intensive ones do not. $U$ is the \emph{internal energy}, $T$ is the
\emph{temperature} and $S$ the \emph{entropy}. For example, consider a mixture
of $r\in\mathbb{N}$ different chemical components. In this case, we have that:

\begin{itemize}
\item $d=r+2$;

\item for $i=1,...,r$, each variable $y_{i}:=N_{i}$ (resp. $-x_{i}:=\mu_{i}$)
represents the number of moles (resp. the chemical potential) of a given type;

\item $x_{r+1}=:P$ is the pressure and $y_{r+1}=:V$ the volume.
\end{itemize}

\begin{remnR}
\label{comp} Sometimes, systems can be seen as composed by \textquotedblleft
simpler\textquotedblright\ ones, i.e. those defined by a smaller number of
variables. We say in this case that such a system is a \emph{composite
system}. In the last example, each chemical compound can be seen as a part of
a composite system.
\end{remnR}

$\ast$ The possible values of the variables $x_{1},...,x_{d-1},T,y_{1}%
,...,y_{d-1},S$ and $U$ give rise to an open manifold $\mathcal{T}%
\subset\mathbb{R}^{2d+1}$ (that we shall assume open), usually called the
\emph{thermodynamical phase space} (TPS): the set of states of the system. We
can see these variables as coordinates for $\mathcal{T}$.

\bigskip

$\ast$ The manifold $\mathcal{T}$ is a \emph{contact manifold} with contact
form%
\begin{equation}
\theta:=dU-T\ dS+%
%TCIMACRO{\dsum \nolimits_{i=1}^{d-1}}%
%BeginExpansion
{\displaystyle\sum\nolimits_{i=1}^{d-1}}
%EndExpansion
x_{i}\,dy_{i}. \label{tita}%
\end{equation}
Thus, $\left(  x_{1},...,x_{d-1},-T,y_{1},...,y_{d-1},S,U\right)  $ defines a
global Darboux system for $\left(  \mathcal{T},\theta\right)  $, see
\cite{Arnold,Mr91}.

\begin{remnR}\label{comp2}
For a composite system (see Remark \ref{comp}) formed out by two simple ones,
the TPS is a product manifold $\mathcal{T}=\mathcal{T}_{1}\times
\mathcal{T}_{2}$ with contact form
\begin{equation}
\theta:=dU_{1}-T_{1}\ dS_{1}+dU_{2}-T_{2}\ dS_{2}+%
%TCIMACRO{\dsum \nolimits_{i=1}^{d_{1}-1}}%
%BeginExpansion
{\displaystyle\sum\nolimits_{i=1}^{d_{1}-1}}
%EndExpansion
x_{1,i}\,dy_{1,i}+%
%TCIMACRO{\dsum \nolimits_{i=1}^{d_{2}-1}}%
%BeginExpansion
{\displaystyle\sum\nolimits_{i=1}^{d_{2}-1}}
%EndExpansion
x_{2,j}\,dy_{2,j}. \label{tita2}%
\end{equation}
Here $\left(  x_{k,1},...,x_{k,d_{1}-1},-T_{k},y_{k,1},...,y_{k,d_{1}-1}%
,S_{k},U_{k}\right)  $, with $k=1,2$, are the global Darboux coordinates of
$\mathcal{T}_{1}$ and $\mathcal{T}_{2}$.
\end{remnR}

$\ast$ By \emph{process }we shall mean every curve $\Gamma:\left[  a,b\right]
\rightarrow\mathcal{T}$. It represents a \textquotedblleft
continuum\textquotedblright\ of actions on the system that produce a
\textquotedblleft continuum\textquotedblright\ of changes on its states.

$\bigskip$

$\ast$ Among the states, a special role is played by a subset $\mathcal{N}%
\subset\mathcal{T}$, known as the \emph{space of equilibrium states}, which is
defined by the following two conditions on $U$. The first one says that, on
the equilibrium states, $U$ and the rest of the extensive variables $y_{i}$'s
and $S$ must be related by the formula
\begin{equation}
U=\Phi\left(  y_{1},...,y_{d-1},S\right)  , \label{U}%
\end{equation}
for some function $\Phi$ (typically homogeneous of degree one). Equation above
is known as the \emph{Fundamental Equation} of the system. The second
condition says that, for any differentiable curve $\Gamma:\left[  a,b\right]
\rightarrow\mathcal{N}\subset\mathcal{T}$, the variation $\Delta U:=U\left(
\Gamma\left(  b\right)  \right)  -U\left(  \Gamma\left(  a\right)  \right)  $
must satisfies
\begin{equation}
\Delta U=\mathbb{Q}-W, \label{deltaU}%
\end{equation}
where $\mathbb{Q}=\mathbb{Q}\left(  \Gamma\right)  $ and $W=W\left(
\Gamma\right)  $ are the heat and the mechanical work, respectively,
interchanged by the system and the environment along the process $\Gamma$.
This is the \emph{First Law of Thermodynamics}.

\begin{rem}
When $\mathbb{Q}=0$ along a process, one says that such a process is
\emph{adiabatic}.
\end{rem}

At a differential level, Eq. (\ref{deltaU}) translates to
\begin{equation}\label{Eq:ConsU}
dU=\dbar\mathbb{Q}-\dbar W.
\end{equation}

Here, $\dbar\mathbb{Q}$ and $\dbar W$ are $1$-forms on $\mathcal{T}$ such that, given a process
$\Gamma$,
\[
\mathbb{Q}\left(  \Gamma\right)  =%
%TCIMACRO{\dint \nolimits_{\Gamma}}%
%BeginExpansion
{\displaystyle\int\nolimits_{\Gamma}}
%EndExpansion
\dbar\mathbb{Q}=%
%TCIMACRO{\dint \nolimits_{a}^{b}}%
%BeginExpansion
{\displaystyle\int\nolimits_{a}^{b}}
%EndExpansion
\left\langle \dbar\mathbb{Q}\left(  \Gamma\left(
t\right)  \right)  ,\frac{d}{dt}\Gamma\left(  t\right)  \right\rangle \,dt.
\]
Idem for $\dbar W$. For instance, for a mixture of
chemical components (see above), $\dbar\mathbb{Q}%
$ and $\dbar W$ are given by
\[
\dbar\mathbb{Q}=T\ dS \qquad \text{and}%
\qquad \dbar W=P\ dV+%
%TCIMACRO{\dsum \nolimits_{i=1}^{r}}%
%BeginExpansion
{\displaystyle\sum\nolimits_{i=1}^{r}}
%EndExpansion
x_{i}\,dy_{i},
\]
at least for some processes. Accordingly,
\[
dU=T\ dS-P\ dV-%
%TCIMACRO{\dsum \nolimits_{i=1}^{r}}%
%BeginExpansion
{\displaystyle\sum\nolimits_{i=1}^{r}}
%EndExpansion
x_{i}\,dy_{i}.
\]
In general, we must have%
\begin{equation}
dU=T\ dS-%
%TCIMACRO{\dsum \nolimits_{i=1}^{d-1}}%
%BeginExpansion
{\displaystyle\sum\nolimits_{i=1}^{d-1}}
%EndExpansion
x_{i}\,dy_{i}. \label{dU}%
\end{equation}
Combining Eqs. (\ref{U}) and (\ref{dU}), it follows that the subset
$\mathcal{N}$ is defined by the equations%
\begin{equation}
x_{i}=-\frac{\partial\Phi}{\partial y_{i}}\left(  y_{1},...,y_{d-1},S\right)
,\ \ T=\frac{\partial\Phi}{\partial S}\left(  y_{1},...,y_{d-1},S\right)
\ \ \ \text{and\ \ \ }U=\Phi\left(  y_{1},...,y_{d-1},S\right)  , \label{ee}%
\end{equation}
known as \emph{state equations}. This means that $\mathcal{N}$ is a
\emph{Legendre submanifold} of $\left(  \mathcal{T},\theta\right)  $ (see
\cite{Arnold,Mr91}).

\begin{remnR}
\label{work}As explained in \cite{Fermi}, the function $\Phi$ or,
equivalently, the internal energy $U$, is defined by the allowed mechanical
work on the system. In other words, $U$ is completely determined if we know
the work done $W\left(  \Gamma\right)  $ along any process $\Gamma$. (This
information, in fact, not only determines $U$, but also $\mathbb{Q}$).\newline
\end{remnR}

$\ast$ For instance, the fundamental equation of the so-called \emph{ideal
gas}, with only one chemical component, is given by$\,$
\begin{equation}
\Phi(N,S,V)=N\,u_{0}\left(  \frac{N\,v_{0}\ e^{\frac{S-N\,s_{0}}{N\,R}}}%
{V}\right)  ^{\frac{1}{\alpha}}, \label{FIG}%
\end{equation}
where $\alpha$ is a dimensionless constant, $R$ is the universal constant of
ideal gases, and $s_{0}$, $v_{0}$ and $u_{0}$ are constants with units of
entropy, volume and energy per mole, respectively (see \cite{Callen} for more
details). Thus, the related state equations read%
\begin{equation}
\mu=\frac{U}{N}\,\left(  1+\frac{1}{\alpha}\,\left(  1-\frac{S}{NR}\right)
\right)  ,\ \ \ T=\frac{U}{NR\alpha},\ \ \ P=\frac{U}{V\,\alpha}, \label{eeig}%
\end{equation}
with%
\begin{equation}
U=N\,u_{0}\left(  \frac{N\,v_{0}\ e^{\frac{S-N\,s_{0}}{N\,R}}}{V}\right)
^{\frac{1}{\alpha}}. \label{feig}%
\end{equation}
Let us mention that the last two equations in (\ref{eeig}) and the Eq.
(\ref{feig}) are usually written as%
\begin{equation}
U=\alpha NRT,\ \ PV=NRT\ \ \ \text{and\ \ \ }S=N\,s_{0}+NR\ \mathrm{ln}\left(
\frac{T^{\alpha}V}{t_{0}^{\alpha}Nv_{0}}\right)  , \label{EEIG}%
\end{equation}
where $t_{0}:=u_{0}/R\alpha$.

\bigskip

$\ast$ The differentiable curves along the equilibrium states $\Gamma:\left[
a,b\right]  \rightarrow\mathcal{N}$ are usually called \emph{quasi-static
processes} (and we shall take this convention). Note that, along such curves,
the state equations (\ref{ee}) are satisfied for every $t\in\left[
a,b\right]  $ (by definition of $\mathcal{N}$).

\begin{remnR}
\label{quasi}In practice, in order to have a quasi-static process $\Gamma$,
the velocity of $\Gamma$ must be small (w.r.t. certain characteristic lengths
and times related to the microscopic properties of the system). That is to
say, the action that defines the process must produce changes in the states at
a very slow rate. This justifies the name \textquotedblleft
quasi-static.\textquotedblright\ However, in this paper, when we say that a
process is quasi-static we will not be assuming that the rate of change of
states is necessarily slow. We will only assume that the state equations are
satisfied for all time along such a process.
\end{remnR}

$\ast$ Not every process $\Gamma:\left[  a,b\right]  \rightarrow\mathcal{N}$
is allowed. According to the \emph{Second Law of Thermodynamics}, a process
$\Gamma$ must satisfy%
\begin{equation}
\Delta S:=S\left(  \Gamma\left(  b\right)  \right)  -S\left(  \Gamma\left(
a\right)  \right)  \geq%
%TCIMACRO{\dint \nolimits_{\Gamma}}%
%BeginExpansion
{\displaystyle\int\nolimits_{\Gamma}}
%EndExpansion
\frac{\dbar\mathbb{Q}}{T}, \label{sl1}%
\end{equation}
where
\begin{equation}%
%TCIMACRO{\dint \nolimits_{\Gamma}}%
%BeginExpansion
{\displaystyle\int\nolimits_{\Gamma}}
%EndExpansion
\frac{\dbar\mathbb{Q}}{T}:=%
%TCIMACRO{\dint \nolimits_{a}^{b}}%
%BeginExpansion
{\displaystyle\int\nolimits_{a}^{b}}
%EndExpansion
\frac{\left\langle \dbar\mathbb{Q}\left(
\Gamma\left(  t\right)  \right)  ,\frac{d}{dt}\Gamma\left(  t\right)
\right\rangle }{T\left(  \Gamma\left(  t\right)  \right)  }\,dt. \label{sl2}%
\end{equation}
In infinitesimal terms%
\begin{equation}
dS\geq\frac{\dbar\mathbb{Q}}{T}. \label{sl3}%
\end{equation}
For adiabatic processes, since $\dbar\mathbb{Q}=0$,
we must have $\Delta S\geq0$.

\bigskip

$\ast$ A process $\Gamma:\left[  a,b\right]  \rightarrow\mathcal{N}$ is say to
be \emph{reversible }if there exists another process $\Gamma^{-}:\left[
a,b\right]  \rightarrow\mathcal{N}$ such that $\Gamma^{-}\left(  a\right)
=\Gamma\left(  b\right)  $ and $\Gamma^{-}\left(  b\right)  =\Gamma\left(
a\right)  $. Otherwise, $\Gamma$ is say to be \emph{irreversible}. Then, a
process is reversible if and only if the equation
\begin{equation}\label{Eq:QE}
\Delta S=%
%TCIMACRO{\dint \nolimits_{\Gamma}}%
%BeginExpansion
{\displaystyle\int\nolimits_{\Gamma}}
%EndExpansion
\frac{\dbar \mathbb{Q}}{T}%
\end{equation}
holds.
\end{appendices}

\bibliographystyle{plain}
\bibliography{bibliografia}
%texto.bib es el fichero donde está salvada la bibliografía.

\end{document}